\documentclass[
  pre,
  reprint,
  superscriptaddress,
  amsmath,amssymb,amsfonts,
  aps
]{revtex4-2}

\usepackage{graphicx}
\usepackage{dcolumn}
\usepackage{bm}
\usepackage{xcolor}
\usepackage[utf8]{inputenc}

\usepackage[caption=false]{subfig}
\usepackage[unicode]{hyperref}
\usepackage[nameinlink,capitalise,noabbrev]{cleveref}

\crefname{subfigure}{panel}{panels}
\Crefname{subfigure}{Panel}{Panels}

\begin{document}

\title{Transition path sampling in Ising models on heterogeneous graphs}

\author{Riccardo Cipolloni}
\affiliation{Dipartimento di Fisica, Sapienza Universit\`a di Roma, Piazzale Aldo Moro 5, 00185 Rome, Italy}

\author{Federico Ricci-Tersenghi}
\affiliation{Dipartimento di Fisica, Sapienza Universit\`a di Roma, Piazzale Aldo Moro 5, 00185 Rome, Italy}
\affiliation{CNR--Nanotec, unità di Roma, Piazzale Aldo Moro 5, 00185 Rome, Italy}
\affiliation{INFN--sezione di Roma 1, Piazzale Aldo Moro 5, 00185 Rome, Italy}

\author{Francesco Zamponi}
\affiliation{Dipartimento di Fisica, Sapienza Universit\`a di Roma, Piazzale Aldo Moro 5, 00185 Rome, Italy}
\date{\today}

\begin{abstract}
Activated transitions have rates that are often exponentially small in system size. Extracting the associated activation barriers is challenging in practice, especially in the deeply metastable regimes and in the presence of disorder. Here, we use transition path sampling to evaluate transition probabilities between ferromagnetic states in the Ising model on finite sparse random graphs, which are perhaps the simplest example of a disordered system with metastable states.
To interpret the transient onset of the transition probability curve, we introduce a minimal three-state kinetic description that highlights the role of intermediate configurations. We validate the method on the heterogeneous Zachary Karate Club network, where distinct dynamical regimes emerge as temperature varies. We then apply the method to random regular graphs and Erd\H{o}s--R\'enyi graphs, showing that sample-to-sample fluctuations are weak in the former but that quenched topological disorder induces sizable instance variability in the latter. For Erd\H{o}s--R\'enyi graphs, we introduce an instance-dependent temperature rescaling that restores a consistent finite-size scaling of dynamical rates and enables a direct comparison with the corresponding static free-energy barrier.
\end{abstract}

\maketitle

\section{Introduction}

A common feature of many-body systems is \emph{multistability}.
This notion has a precise dynamical meaning: on experimentally or numerically accessible time scales, the system can admit several long-lived macrostates, with transitions between them strongly suppressed by large activation barriers.
The paradigm is already provided by the ferromagnetic Ising model below its critical temperature, where the two symmetry-related magnetized phases behave as metastable basins under local dynamics; more generally, multistability arises whenever collective degrees of freedom are effectively confined near distinct attractors of the coarse-grained dynamics. In such settings, relaxation is naturally \emph{intermittent}: the system typically spends long times fluctuating within one macrostate (where local equilibration can take place) and only rarely performs a barrier-crossing event that relocates it to another macrostate. These activated switches control the long-time kinetics of phase changes and switching phenomena across a wide range of contexts, from nucleation-dominated transformations in magnetic systems to collective rearrangements in heterogeneous and complex media~\cite{hangii_review}. A central goal is therefore to \emph{quantify} the inter-state transition rates and their dependence on control parameters (e.g.\ temperature) and on system size, because these rates provide a quantitative link between microscopic stochastic dynamics and macroscopic long-time behavior. Doing so, however, is difficult: genuinely activated transitions are so rare that straightforward dynamical sampling is impractical, and the resulting estimates can be dominated by finite-time effects and by the operational definition of macrostates.

In general, one expects the transition rate to become exponentially small with the number of degrees of freedom involved in the transition process.
In finite-dimensional Ising models, the transition requires forming an interface, and the free energy cost scales as a surface. In mean-field systems, either fully connected or defined on a random graph geometry, a finite fraction of the degrees of freedom is involved,
reflecting an extensive free energy barrier cost. 
A convenient finite-size statement of this expectation is the Arrhenius form for the rate
\begin{equation}
\label{eq:ArrhLaw}
  k(\beta) \;\simeq\; \tilde{k}(\beta)\,\exp\!\bigl[-N\,\beta\,\Delta f(\beta)\bigr],
\end{equation}
where $\beta$ is the inverse temperature, $\Delta f(\beta)$ is an intensive free energy barrier cost, and $\tilde{k}(\beta)$ is a prefactor that is at most sub-exponential in $N$.
Eq.~\eqref{eq:ArrhLaw} makes explicit why rate measurements are hard: extracting $\Delta f(\beta)$ requires resolving an exponentially small rate while controlling protocol-dependent effects such as transient times and the operational definition of macrostates.

A common route to rate calculations is to identify a transition state along a low-dimensional order parameter and then estimate the rate from trajectories initiated near that dividing surface. For complex high-dimensional systems, however, transition states are often not known or not uniquely specifiable, and widely used order parameters can be poor approximations to the true reaction coordinate~\cite{Bolhuis02, Dellago98}. This motivates rare-event approaches that avoid brute-force time sampling by working instead with ensembles of reactive trajectories, or by estimating probability fluxes between metastable sets~\cite{valerianirev}. Representative alternatives include interface-based schemes such as forward-flux sampling~\cite{tenwoldeffs} and trajectory-selection methods based on population dynamics~\cite{kurchan1,kurchan2}. A different possibility is to encode the target event in a modified dynamics, provided that the required conditioning weights are accessible. This yields bridge dynamics through a modified generator~\cite{majumdarorland2015,aguilar2022}; in complex systems, though, such constructions typically require reduced descriptions or small-noise optimal-control approximations~\cite{vandenweare2012,mamedecoghi2026}.

Among trajectory-space approaches, Transition Path Sampling (TPS) constructs a Monte Carlo sampling directly on dynamical trajectories and thereby generates representative reactive paths conditioned on prescribed initial and final basins, without requiring an explicit reaction coordinate~\cite{Dellago98, Bolhuis02}. In the setting of discrete interacting variables, a particularly convenient implementation adapts TPS to local updates of single-variable paths and can be combined with thermodynamic integration in an appropriate control parameter to reconstruct constrained transition probabilities and infer rates effectively~\cite{ZMW_2012}.

When heterogeneity is \emph{quenched} into the system, activated kinetics acquires an additional layer of complexity: the transition rate is no longer a single quantity to be measured at given $(N,\beta)$, but becomes a sample-dependent observable that fluctuates across realizations of the disorder. This is especially acute in the activated regime because rates are exponentially sensitive to the effective barrier.
In homogeneous settings, the dominant reactive mechanism is often essentially unique up to trivial symmetries (e.g.\ translations of a critical nucleus), so that different transition paths largely represent shifted versions of the same elementary event. Under quenched heterogeneity, by contrast, distinct mesoscopic ``channels'' can compete: barrier crossings may preferentially involve different locations or topological motifs, yielding families of reactive trajectories that are not related by simple symmetries.
In this situation, relatively small differences in local environments or mesoscopic pathways can translate into large variations of $k(\beta)$, and hence into large sample-to-sample fluctuations at fixed size and temperature.
In practice, one is then faced with a distribution of rates (or of free energy barriers) rather than a single one, and the approach to self-averaging can be slow on the system sizes accessible to numerical work~\cite{random_graphs_review,disorder_dynamics_review}. 

A simple example where these features naturally appear is disordered networks and sparse graphs, where local connectivity and interaction neighborhoods fluctuate across nodes and across graph instances. Already in this relatively simple setting, extracting an intensive barrier function $\Delta f(\beta)$ from dynamical measurements is not only a matter of overcoming the rarity of transitions, but also of designing an operational procedure that remains stable under disorder fluctuations and yields a coherent finite-size scaling across samples. Studying such problems is the first important step towards more complex systems, because sparse-graph ferromagnets provide a standard testing ground where sharp static predictions for free-energy barriers can be obtained within the Bethe (or cavity) method, while the corresponding dynamical extraction is precisely where quenched variability and finite-time effects are most severe. Developing and validating a disorder-robust dynamical route to $\Delta f(\beta)$---and assessing when and how it can be brought into quantitative contact with static Bethe barriers---is the specific gap addressed in the present work.

In this work, we study activated transitions in the ferromagnetic Ising model on finite sparse graphs and assess how quenched topological disorder impacts the dynamical estimation of rates and finite-size barriers. Building on the Transition Path Sampling (TPS) and thermodynamic-integration framework of Ref.~\cite{ZMW_2012}, we first summarize the path-sampling protocol (Sec.~\ref{sec:algorithm}) and the subsequent thermodynamic integration in inverse temperature. The central output is the constrained transition probability $Z_{X,Y}(T;\beta)$, i.e.\ the probability that a trajectory started in macrostate $X$ ends in macrostate $Y$ at time $T$ and inverse temperature $\beta$, reconstructed by integrating $U(T,\beta)=\partial_\beta \ln Z_{X,Y}(T;\beta)$ along a $\beta$-ladder. We then show how a single long-$T$ dataset can be exploited to obtain $Z_{X,Y}(t)$ for $t\le T$ and, whenever a linear--growth window is present, to extract an effective transition rate from its growth (Sec.~\ref{ssec:samplingTrajectories}). To interpret the time dependence of $Z_{X,Y}(t)$ in metastable regimes and to make the rate-extraction procedure mechanistically transparent, we introduce an effective three-state kinetic picture $X\to I\to Y$ that isolates intermediate-state trapping through residence times and provides a minimal mechanism for the emergence of a linear--growth regime in $Z_{X,Y}(t)$ (Sec.~\ref{sec:3StatesModel}).

We next use the Zachary Karate Club network as a heterogeneous benchmark to illustrate the resulting diagnostics across temperatures and to validate the operational conditions under which the time-reweighting reconstruction is reliable (Sec.~\ref{sec:ZKC}). Finally, we apply the pipeline to random-graph ensembles: for random regular graphs we extract size-dependent barriers and compare them to the corresponding static Bethe prediction (Sec.~\ref{sec:RRG}); for Erd\H{o}s--R\'enyi graphs we address strong instance-to-instance variability at fixed $(N,\beta)$ by introducing an instance-dependent inverse-temperature rescaling that realigns transition data across graphs and restores a coherent size dependence, enabling an ensemble-level barrier extraction and a comparison with the Bethe prediction (Sec.~\ref{sec:ER}). Details are collected in the Appendix.

\section{Path-Sampling Algorithm}
\label{sec:algorithm}

In this section, we summarize the Monte Carlo (MC) strategy used to sample rare transition paths according to the procedure developed in
Ref.~\cite{ZMW_2012}, highlighting the minor modifications introduced for the present work. After fixing notation and dynamical rates for the single variables (Sec.~\ref{ssec:setting}), we recall the thermodynamic integration (TI) procedure that yields the path–ensemble probability, and we explain how a single TI carried out at a sufficiently long trajectory time \(T\) can be reweighted to reconstruct the full transition-probability curve and extract transition statistics such as the operational rate (Sec.~\ref{ssec:samplingTrajectories}).
A more detailed description of the procedure is presented in Ref.~\cite{ZMW_2012}.

\subsection{Setting and notation}
\label{ssec:setting}

We consider an Ising‐like system, consisting of $N$ binary spins
$s_i(t)\!=\!\pm1$ evolving under continuous‐time single-spin
Glauber dynamics.
For a generic configuration $\underline{s}$ we define
\begin{eqnarray}
  H(\underline{s}) \;=&&\; -\sum_{i<j} J_{ij}\,s_i s_j 
  \;-\; \sum_{i} h_i^{\text{ext}}\,s_i ,
  \\
  h_i(\underline{s}) \;\equiv&&\;
      h_i^{\text{ext}} + \sum_{j(\neq i)} J_{ij}\,s_j ,
\end{eqnarray}
where $\boldsymbol{J}$ is the interaction matrix and $\underline{h}^{\text{ext}}$ is a site-dependent external field,
so that flipping spin $i$ costs an energy
$\Delta E_i = 2s_i h_i$.  
Following Ref.~\cite{ZMW_2012} we choose \emph{symmetric} spin-flip
rates (we set Boltzmann constant $k_\mathrm{B}=1$; hence $\beta=1/\mathcal{T}$ for a temperature $\mathcal{T}$)
\begin{equation}
\label{eq:rates}
  w_{s_i\to-s_i} \;=\;
  e^{-\beta\,\Delta E_i/2},
\end{equation}
which obey detailed balance.

A trajectory from time $t=0$ to time $t=T$ is the full time record
$\underline{s}(t)\equiv\{s_i(t)\}_{i=1}^N$ on $t\!\in\![0,T]$. A complete path is therefore weighted by an \emph{initial} Boltzmann
factor $P_{\mathrm X}[\underline{s}(0)]$ and by a \emph{final} projector
(or bias) $\chi_{\mathrm Y}[\underline{s}(T)]$, whose precise forms are
chosen to pin the trajectory in the desired macrostates at $t\!=\!0$ and
$t\!=\!T$. If the macrostates are defined by the
instantaneous (extensive) magnetization $M(t)=\sum_i s_i(t)$, as in ferromagnetic models, a convenient choice is
\begin{eqnarray}
  P_\text{X}[\underline{s}(0)]
  \;=&&\;
  \frac{\exp\!\bigl[h_{in}\,M(0)\bigr]}%
       {Z(h_{in})},
  \\
  \chi_\text{Y}[\underline{s}(T)]
  \;=&&\;
  \exp\!\bigl[h_{out}\,\Theta(M(T)-M^\star)\bigr],
  \label{eq:BC_general}
\end{eqnarray}
$M^\star$ being a prescribed extensive threshold, and denoting the corresponding intensive threshold by $m^\star \equiv M^\star / N$.
$Z(h_{in})$ normalizes the initial Boltzmann weight.
$\Theta$ is the Heaviside step function with the convention $\Theta(0)=1$.

With finite values for $h_{in}$ and $h_{out}$ one can explore a continuum of
initial and final biases, recovering the unbiased ensemble by setting
$h_{in}=h_{out}=0$.
If the macrostates are such that their internal relaxation times are much shorter than the observational scale, and that their structure is robust enough --- as in the ferromagnetic cases at low-temperature discussed in this work --- any large-enough value for $h_{in}$ and $h_{out}$ is effective in selecting the initial and final states.
We hereby work in the hard-wall limit
\begin{equation}
  h_{in}\to -\infty,
  \qquad
  h_{out}\to +\infty ,
\end{equation}
which pins the initial configuration in the fully polarised $(-)$ state and
selects trajectories whose endpoint lies beyond the threshold $M^\star$.
More precisely, up to a trajectory-independent factor in $\chi_Y$,
\begin{eqnarray}
  P_\text{X}[\underline{s}(0)]
  \;\longrightarrow&&\;
  \delta_{\,M(0),-N},
  \nonumber\\
  \chi_\text{Y}[\underline{s}(T)]
  \;\longrightarrow&&\;
  \Theta\!\bigl(M(T)-M^\star\bigr).
  \label{eq:BC_limit}
\end{eqnarray}
Equivalently, in terms of the intensive magnetisation $m(t)=M(t)/N$, one has
$P_\text{X}\to \delta_{\,m(0),-1}$ and $\chi_\text{Y}\to \Theta(m(T)-m^\star)$

The prescribed $M^\star$ must fulfill two requirements: 
\begin{itemize}
    \item it must lie \emph{above} any magnetization attainable without crossing the nucleation barrier out of the $(-)$ macrostate, so that reaching $M^\star$ necessarily signals the required crossing event; 
    \item it must stay \emph{below} the lowest magnetization typically reached during \emph{intra-state fluctuations} inside the $(+)$ well, so as not to discard legitimate escape paths that have already entered the final macrostate but have not yet attained perfect alignment.
\end{itemize}

We find empirically that \(m^\star\simeq 0.6\) is a choice that matches these requirements for both the random graphs cases discussed in this paper, when $\beta=1$. 
At higher, still-ferromagnetic temperatures, intra-state fluctuations become larger.
Although a different $M^\star$ would in principle be optimal at higher (still ferromagnetic) temperature, we keep it fixed throughout this work to make the definition of the final-state projector $\chi_Y$ uniform across samples and across the $\beta$-ladder used in TI. When the choice of $M^\star$ induces a measurable bias (due to post-entry fluctuations), we correct it a posteriori in the rate extraction whenever possible (see Appendix~\ref{sec:app_moderateBeta}).

Eq.~\eqref{eq:BC_limit} fixes the trajectory to start deep inside
the $X$ macrostate ($m=-1$) and \emph{only} requires the end‐point to exceed
the threshold $m^\star$.
We have numerically verified that this hard-wall choice yields the same
escape rate as the finite-fields formulation of
Eq.~\eqref{eq:BC_general} in the expected compatible scenarios.

\subsection{The sampling-and-integration protocol}
\label{ssec:samplingTrajectories}

We now outline the essential steps of the path-sampling MC.
More details can be found in Ref.~\cite{ZMW_2012};
here we only re-state the ingredients strictly needed to reproduce our
runs.

\subsubsection{Trajectory initialization}

One needs \emph{one} complete path compatible with the hard boundary
conditions in Eq.~\eqref{eq:BC_limit} in order to start the path-sampling Markov chain.
As a convenient seed we use the minimal ``template'' used in Ref.~\cite{ZMW_2012}:
initial spins fixed as $s_i=-1$, final spins fixed as $s_i=1$, and a single flipping-time drawn independently from a uniform distribution over $(0,T)$ for each spin.
Starting from this seed, we use two operational initialization protocols:
(i)~a \emph{direct start}, where the chain is equilibrated directly at the target inverse temperature $\beta_{\mathrm{tgt}}$ by single-spin path updates as in Ref.~\cite{ZMW_2012}, and measurements are taken only after convergence; and
(ii)~an annealed initialization (\textsc{anneal-init}), in which the path ensemble is progressively cooled from $\beta=0$ to $\beta_{\mathrm{tgt}}$ while keeping the endpoint constraints fixed.
During \textsc{anneal-init} we temporarily enforce the final constraint as $m^\star=1$ (equivalently, we keep $h_{in}\to-\infty$ and $h_{out}\to+\infty$ and set $m^\star=1$), and once $\beta_{\mathrm{tgt}}$ is reached we switch back to the working threshold $m^\star$ for subsequent sampling.
In practice, we use a uniform $\beta$ ladder and increase the number of sweeps performed at each fixed $\beta$ geometrically.
For large and/or highly heterogeneous graphs, \textsc{anneal-init} substantially reduces the equilibration time.
Whenever we performed matched runs differing only in the initialization protocol, we found no systematic differences in observables or fitted kinetic parameters after convergence (within statistical and fit-procedure variability).

\subsubsection{Single-spin path update}

The code updates one spin $i$, uniformly chosen, at each step.
Its neighbors' histories remain fixed, so the local field $h_i(t)$ is piecewise-constant; the interval boundaries $t_0=0<t_1<\cdots<t_K=T$ are determined by flips of first- and second-neighbor spins.
On each interval $(t_{k-1},t_k)$ the unnormalized conditional propagator of $s_i(t)$ is the $2\times2$ matrix $\bigl[e^{\lambda\widehat L_i}\bigr]_{s',s}$, with $\lambda=t_k-t_{k-1}$ and $\widehat L_i$ the conditional operator defined in Ref.~\cite{ZMW_2012}.
Given the sampled interval boundary values, diagonalizing $\widehat L_i$ once gives the \emph{exact} probability that $s_i$ (a)~stays unchanged on the interval and (b)~flips after a waiting time $u<t_k-t_{k-1}$.
Waiting times are drawn until $t=T$; the fresh $s_i(t)$ is then
accepted unconditionally.  
Apart from this concise summary, we refer to
Ref.~\cite{ZMW_2012} for the explicit expressions.

\subsubsection{Thermodynamic integration}
\label{ssec:TI}

Denote by
\begin{equation}
  Z_{X,Y}(T;\beta)=\Bigl\langle \chi_Y[\underline s(T)] \Bigr\rangle_{X,T,\beta}
\end{equation}
the normalization (``constrained partition function'') of the path ensemble with initial weight $P_X[\underline s(0)]$, i.e.\ the average of $\chi_Y[\underline s(T)]$ over unbiased trajectories of duration $T$ started from $P_X$. 
Its logarithmic derivative,
\begin{equation}
   U(T,\beta)
  =\frac{\partial}{\partial\beta}\ln Z_{X,Y}(T;\beta)
  =\bigl\langle \mathcal U[\underline s(t)]\bigr\rangle_{X,Y,T,\beta} \ ,
\label{eq:U_def}
\end{equation}
can be computed over the sampled paths ensemble.  Writing the trajectory as
intervals of constant configuration $\{(t_{k-1},t_k)\}$ one has
\begin{equation}
 \mathcal U[\underline s(t)]
  =
  \sum_{k=1}^{K}
  \int_{t_{k-1}}^{t_k}\!\!dt\,
      \sum_{i} w_{s_i\to -s_i}\,s_i h_i
  - \sum_{k=1}^{K-1} h_{i_k}s_{i_k},
\end{equation}
where $i_k$ is the spin that flips at $t_k$.
Sampling $ U(T,\beta)$ on a ladder
$0=\beta_0<\dots<\beta_m=\beta^\star$ and spline-interpolating
it, the constrained weight at $\beta_{\mathrm{tgt}}$ follows from
\begin{equation}
  \ln Z_{X,Y}(T;\beta_{\mathrm{tgt}})
  = \ln Z_{X,Y}(T;0)
    +\int_0^{\beta_{\mathrm{tgt}}}\! U(T,\beta)\,d\beta .
\end{equation}
At high temperature (small $\beta$), for the trajectory horizons $T$ used in practice, the configuration-space relaxation time is shorter than $T$ and finite-time effects on $U(T,\beta)$ become negligible within statistical fluctuations.
In this regime, $U(T,\beta)$ is numerically indistinguishable from its $T\!\to\!\infty$ limit, which can be estimated efficiently by standard equilibrium Monte Carlo (see Ref.~\cite{ZMW_2012} and  Appendix~\ref{ssec:app_TI_procedure}). We therefore anchor the low-$\beta$ part of the TI integrand with the standard-MC estimator and switch to TPS only beyond a matching point where the two estimators start to deviate appreciably.

\subsubsection{Rate extraction from a single TI}
\label{ssec:rateTrick}

In line with the TPS rate-calculation framework~\cite{Dellago98}, and with the practical strategy used in Ref.~\cite{ZMW_2012}, we reconstruct $Z_{X,Y}(\tau)$ from a single TI at horizon $T$ via a time-reweighting relation.

Indeed, the probability of finding the system in \(Y\) at an earlier time \(\tau<T\) can be estimated by reweighting on the sampled paths ensemble
\begin{equation}
  Z_{X,Y}(\tau)=
  Z_{X,Y}(T)\,
  \bigl\langle \chi_Y[\underline s(\tau)] \bigr\rangle_{X,Y,T, \beta} \ ,
  \label{eq:reweight}
\end{equation}
if the condition that the jump occurs only once within the chosen time window, 
\begin{equation}
  \mathbb P\!\bigl(
      \chi_Y[\underline s(t')]=1 \,\big|\, \chi_Y[\underline s(t)]=1
    \bigr)\simeq 1,
  \qquad\forall\,T>t'>t
  \label{eq:persistence}
\end{equation}
is satisfied in general, i.e., also out of the sampled ensemble generated via the conditioning on the configurations at time $T$.

Studying the transition in proper $2$-state systems, if Eq.~\eqref{eq:persistence} holds and $T$ exceeds the transient time, the curve $Z_{X,Y}(\tau)$ reconstructed from Eq.~\eqref{eq:reweight} faithfully reflects the transition history and can be analysed to extract relevant parameters, e.g., transition rates.
In more complex scenarios where many diverse transitions among states are considered, if Eq.~\eqref{eq:persistence} holds the reweighting trick is still a viable way to reconstruct the probability over time for being in the selected final state. Determining a priori whether Eq.~\eqref{eq:persistence} holds outside the sampled, constrained ensemble is, however, in general not possible.

The persistence requirement in Eq.~\eqref{eq:persistence} typically breaks down for two reasons:
\begin{itemize}\setlength\itemsep{0.2em}
  \item \emph{Post-transition fluctuations.}  
        The threshold $m^\star$ is set high enough to certify that the trajectory has \emph{reached} the final macrostate, yet the broader equilibrium distribution in $Y$ allows the magnetisation to dip below that value for short intervals, so that $\chi_Y$ intermittently returns to~0 and $Z_{X,Y}(\tau)$ is underestimated.
  \item \emph{Transient excursions.}  
        At moderate or high temperature, or in rough landscapes, the system may leave $Y$ one or more times before~$T$, invalidating the single-pass picture implicit in Eq.~\eqref{eq:persistence}.
\end{itemize}

If such departures are brief and infrequent, one can correct the inferred rate either by applying the explicit threshold-recrossing correction described in Appendix~\ref{sec:app_moderateBeta},
or by evaluating the averages with a softer projector $\tilde\chi_Y$ (defined with a less demanding threshold) on the \emph{same} sampled paths.
When excursions become long or frequent, the reweighting formula in Eq.~\eqref{eq:reweight} ceases to be reliable and independent TI runs at several trajectory lengths~$T$ are required.

\subsubsection{Code flow}

Operatively, the algorithm can be cast as follows.
\begin{enumerate}
  \item Fix the macrostate projectors ($h_{in}$, $h_{out}$, $m^\star$), a set of trajectory horizons $\{T_\alpha\}$, and a $\beta$-ladder $\{0=\beta_0<\dots<\beta_m=\beta^\star\}$ up to a maximum inverse temperature $\beta^\star$.
  \item For each chosen horizon $T_\alpha$:
    \begin{enumerate}
      \item Initialise one trajectory compatible with the hard boundary conditions at $T_\alpha$ (optionally via \textsc{anneal-init}).
      \item For each $\beta_j$ on the ladder, run TPS at fixed $(T_\alpha,\beta_j)$ until convergence of path observables and measure $ U(T_\alpha,\beta_j)$ (anchoring the low-$\beta$ part with the standard-MC estimator as described above and shown in Appendix \ref{ssec:app_TI_procedure}).
      \item Integrate $ U(T_\alpha,\beta)$ along the $\beta$-ladder to obtain $Z_{X,Y}(T_\alpha;\beta)$ on the ladder.
    \end{enumerate}
  \item For the $\beta$ values of interest on the ladder, construct the transition-probability curve $Z_{X,Y}(t)$ over the time range of interest:
    \begin{enumerate}
      \item either by time-reweighting as in Eq.~\eqref{eq:reweight} from a single sufficiently long horizon $T$ when Eq.~\eqref{eq:persistence} is satisfied, or
      \item by combining independent estimates $\{Z_{X,Y}(T_\alpha;\beta)\}$ obtained at several horizons $T_\alpha$.
    \end{enumerate}
  \item Analyse $Z_{X,Y}(t)$ and fit it to extract kinetic parameters (e.g.\ an operational rate $k_{X\to Y}$ can be obtained whenever a linear-growth regime is observed). Uncertainty estimates we report in the following refer to the fitting procedure. Additional contributions stem from the thermodynamic integration, from the statistical uncertainty in the reconstructed curve \(\chi(t)\), and from the determination of the linear-growth fitting window. The first contribution is estimated in Appendix~\ref{ssec:app_TI_errors} by a blocking/bootstrap analysis of the TI integrand, while the other two have been monitored operationally; the estimated overall effect remains small across the instances and models checked, and is too small to affect the subsequent analysis.
\end{enumerate}

For most of the transition rates probed in the following, direct dynamics estimates would be effectively out of reach; in Appendix~\ref{sec:app_compAdv} we provide an order-of-magnitude estimate of the computational advantage of the TPS+TI strategy on one of the hardest instances.

\section{Three-State Model for Metastable Transitions}
\label{sec:3StatesModel}

A three-state model can provide a refined description of metastable transitions by introducing an intermediate state that accounts for partial reconfigurations before a full transition occurs. This section expands on the mathematical structure and the physical significance of this formulation.

\subsection{Definitions}

Consider a system whose state $S$ can take $3$ values:
\begin{itemize}
    \item \textbf{$X$}: The initial state, in which the system remains until a fluctuation induces a transition.
    \item \textbf{$I$}: An intermediate state where the system has undergone partial reconfiguration, forming nucleation centers or local clusters.
    \item \textbf{$Y$}: The target state for our study, determining the completion of the transition.
\end{itemize}

We model $S(t)$ as a continuous-time Markov chain with constant transition rates $k_{S\to S'}$, i.e.\ exponentially distributed waiting times in each state.
Then, the state probabilities $P_A(t)\equiv\mathbb P(S(t)=A)$ evolve according to:
\begin{equation}
\label{eq:3StateModelODE}
\left\{ 
\begin{aligned} 
    \frac{dP_X}{dt} &= -k_{X \to I} P_X + k_{I \to X} P_I, \\ 
    \frac{dP_I}{dt} &= k_{X \to I} P_X - (k_{I \to X} + k_{I \to Y}) P_I + k_{Y \to I} P_Y, \\ 
    \frac{dP_Y}{dt} &= k_{I \to Y} P_I - k_{Y \to I} P_Y,
\end{aligned} 
\right. 
\end{equation}
with initial condition $P_S(0)=\delta_{S,X}$.
Note that here we assume that the transition must necessarily go through the intermediate state $I$, as no direct processes $X \to Y$ and $Y\to X$ are taken into account. Hence, the case of three mutually equivalent states --- for general time-scales --- cannot be described by Eq.~\eqref{eq:3StateModelODE}.

The state $I$ can also be viewed as a collective label for $\mathcal{N}$ microstates $\{I_i\}$, each having the same (or negligibly different) exiting rates $k_{I_i \to X}, k_{I_i \to Y}$. In this case the state $I$ would be such that  $P_I=\sum_{i=1}^{\mathcal{N}}P_{I_i}$, $k_{X,Y\to I}=\sum_{i=1}^{\mathcal{N}} k_{X,Y\to I_i}$, $k_{I\to X,Y}=\sum_{i=1}^{\mathcal{N}} P_{I_i} k_{I_i \to X,Y} / P_I$.
If the exit rates are only approximately equal, this reduction is meant as an effective coarse-graining, accurate when intra-$I$ relaxation is fast compared to escapes.

\subsection{Short-Time Behavior: Quadratic Growth}
\label{subsec:QuadraticGrowth}
Let $k_{\rm out}^{\max}\equiv \max\{k_{X\to I},\,k_{I\to X}+k_{I\to Y},\,k_{Y\to I}\}$.
For times $t\ll (k_{\rm out}^{\max})^{-1}$ one has the universal short-time expansion
\begin{equation}
\label{eq:quadraticOnset}
    P_Y(t) \approx \frac{k_{X \to I} k_{I \to Y}}{2} t^2 + O(t^3) \ ,
\end{equation}
with a leading coefficient that only involves the two-step pathway $X\to I\to Y$.
If the \emph{backward} channels $I\to X$ and $Y\to I$ can be neglected on the observation window, then one effectively has
$k_{\rm out}^{\max}=\max\{k_{X\to I},\,k_{I\to Y}\}$.

Also in this regime, conditioning on $S(T)=Y$ selects, to leading order in $k_{\rm out}^{\max}T$,
trajectories involving exactly two jumps $X\to I\to Y$ and no further transitions up to time $T$.
Let $\tau_1$ be the sojourn time in $X$ before the first jump and $\tau_2$ the subsequent sojourn time in $I$ before the jump to $Y$
(hence $\tau_1,\tau_2\ge 0$ and $\tau_1+\tau_2\le T$).
At leading order one may drop all factors of order $O(k_{\rm out}^{\max}T)$, obtaining the joint density of $(\tau_1,\tau_2)$ together with the event $S(T)=Y$:
\begin{eqnarray}
\label{eq:triangle_unnorm}
  p_{\tau_1,\tau_2}\bigl(x,y;\,S(T)&&\!\!\!\!\!=Y\bigr)
  \simeq\\
  \nonumber
  &&k_{X\to I}\,k_{I\to Y}\;
  \mathbf 1_{\{x\ge0,\,y\ge0,\,x+y\le T\}} \ ,
\end{eqnarray}
where corrections come from trajectories with extra jumps and are of higher order in $k_{\rm out}^{\max}T$.
Using Eq.~\eqref{eq:quadraticOnset} at $t=T$,
the conditional p.d.f.\ reads
\begin{equation}
\label{eq:triangle_uniform}
  p_{\tau_1,\tau_2}\bigl(x,y \mid S(T)=Y\bigr)
  \;\simeq\;
  \frac{2}{T^2}\;
  \mathbf 1_{\{x\ge0,\,y\ge0,\,x+y\le T\}} \ .
\end{equation}
This implies an equipartition of the observation time among the three sojourns in $X$, $I$, and $Y$:
\begin{eqnarray}
\mathbb{E}\bigl[\tau_{1}\mid && \!\!\!\!\! S(T)=Y\bigr]
=\,
\mathbb{E}\bigl[\tau_{2}\mid S(T)=Y\bigr]
\nonumber
\\
=&&
\!\!\!\!\!
\mathbb{E} \bigl[T-(\tau_{1}+\tau_{2})\mid S(T)=Y\bigr]
=\frac{T}{3},
\end{eqnarray}
with $T-(\tau_1+\tau_2)$ the permanence time in $Y$.

\subsection{Exact Dynamics Overview}
\label{ssec:exactGeneral}

The general solution of Eq.~\eqref{eq:3StateModelODE} is a superposition of two nonzero relaxation modes and the steady–state:
\begin{equation}
  P_Y(t)
  \;=\;P_Y(\infty)
     \;-\;C_1e^{-\lambda_1t}
     \;-\;C_2e^{-\lambda_2t},
\end{equation}
with $C_1+C_2=P_Y(\infty)$ under $P_Y(0)=0$.  Only when these rates are widely separated, $\lambda_2\gg\lambda_1$, does the faster mode drop out within a certain observational window, leaving an approximately linear growth
\begin{equation}
  P_Y(t)
  \simeq P_Y(\infty) -C_1+ C_1\,\lambda_1\,t
  \qquad \text{for}\quad \lambda_2^{-1}\ll t\ll \lambda_1^{-1}.
\end{equation}
By contrast, when the nonzero eigenvalues are comparable—as in structured graphs and in glassy landscapes with competing intermediate-state escape and return rates---no parametrically clean linear--growth window appears, and the growth of $P_Y(t)$ retains pronounced slowing down even for $P_Y$ very small, i.e., much smaller than its asymptotic value.

As a concrete illustration, consider the extreme “all‐rates‐equal” limit
\begin{equation}
  k_{X\to I}=k_{I\to X}=k_{I\to Y}=k_{Y\to I}=k.
\end{equation}
The corresponding rate matrix has eigenvalues $\{0,\,k,\,3k\}$, and one finds the exact solution
\begin{equation}
  P_Y(t)
  =\frac{1}{3}
   \;-\;\frac12\,e^{-kt}
   \;+\;\frac16\,e^{-3kt}
  \;=\;\frac12\,k^2\,t^2 + O(t^3).
\end{equation}
The two nonzero modes differ only by a factor of three, so neither exponential can be peeled away on any intermediate time window.  One therefore sees only a smooth crossover from the initial quadratic growth into the eventual saturation, all governed by the single scale $k^{-1}$.

\subsection{Removing the Intermediate State}
\label{ssec:quasiStaticI}

When the intermediate state $I$ equilibrates fast with respect to the overall transition characteristic times, one may set
$\dot P_I\simeq0$ in the three‐state ODEs, Eq.~\eqref{eq:3StateModelODE}, yielding
\begin{equation}
  P_I(t)\;\approx\;
  \frac{k_{X\to I}\,P_X(t) + k_{Y\to I}\,P_Y(t)}{k_{I\to X}+k_{I\to Y}}.
\end{equation}
Substituting back gives an effective two‐state kinetics
\begin{equation}
  \dot P_X = -k_{\rm eff}P_X + k'_{\rm eff}P_Y,
  \quad
  \dot P_Y =  k_{\rm eff}P_X - k'_{\rm eff}P_Y,
\end{equation}
with
\begin{equation}
  k_{\rm eff}
  = \frac{k_{X\to I}\,k_{I\to Y}}{k_{I\to X}+k_{I\to Y}},
  \quad
  k'_{\rm eff}
  = \frac{k_{Y\to I}\,k_{I\to X}}{k_{I\to X}+k_{I\to Y}}.
\end{equation}

For the genuine three-state initial condition
$P_X(0)=1$, $P_I(0)=P_Y(0)=0$, the quasi-static reduction applies only after
the fast relaxation of $I$, on a time scale
\begin{equation}
  \tau \;\simeq\; (k_{I\to X}+k_{I\to Y})^{-1}.
\end{equation}
Matching the post-transient dynamics to the effective two-state kinetics then yields an approximately linear growth
\begin{equation}
  P_Y(t)\;\approx\;k_{\rm eff}\,\bigl(t-\tau\bigr)
  \qquad
  \text{for}\;
  \tau\ll t\ll (k_{\rm eff}+k'_{\rm eff})^{-1}.
\end{equation}
The renormalized rate $k_{\rm eff}$ encodes the interplay between metastable excitations and their dynamical pathways. This derivation also highlights how the transient time $\tau$ depends on the time-scales of microscopic and mesoscopic motion, and hence on tuning parameters like the temperature.

In a standard Ising ferromagnet, the two-state description considers only the transitions between the states $(-)$ and $(+)$. However, the actual transition process involves intermediate ``nucleated'' states, which must be reached before the system completes the flip. Considering a single such state $I$ --- for instance, because it is symmetric under spin-flip in terms of dynamic accessibility --- this introduces an effective transition rate given by:
\begin{equation}
    k_{+ \to -} = \frac{k_{+ \to I} k_{I \to -}}{k_{I \to +} + k_{I \to -}}.
\end{equation}
Also, for a system where there are more, distinct and independent pathways from $X$ to $Y$, each passing through an intermediate state, $I_i$, and displaying a transient time $\tau_i$, if all the $\tau_i$ are much smaller than the equilibration time-scales for $X$ and $Y$ we obtain the $P_Y(t)$ behaviour just by summing all pathways contributions. For instance, we have a quadratic behaviour $\propto \sum_i \tfrac12 k_{X\to I_i} k_{I_i \to Y} t^2$ for $t\ll \min(\{\tau_i\})$, and a late-time linear behaviour $\propto \sum_i k_{\rm eff}^{(i)} t$ for $\max_i \tau_i \ll t \ll \min_i (k_{\rm eff}^{(i)})^{-1}$ (assuming negligible backflow from $Y$ to the intermediates), with
$ k_{\rm eff}^{(i)}=\tfrac{k_{ X\to I_i } k_{ I_i \to Y}}{k_{ I_i \to X}+ k_{ I_i \to Y}}$.

Considering, as an example, the case of an Ising ferromagnet in its low-temperature phase displaying 2 states, $X\!=\!(+)$ and $Y\!=\!(-)$, in the absence of external fields, we have
\begin{equation}
  k_{X\to I}=k_{Y\to I}=k,
  \quad
  k_{I\to X}=k_{I\to Y}=r\,k,
  \quad
  r\gg1.
\end{equation}
Then
\begin{equation}
  k_{\rm eff}=\frac{k\,(rk)}{2rk}=\frac{k}{2}, 
  \quad
  k'_{\rm eff}=\frac{k\,(rk)}{2rk}=\frac{k}{2},
\end{equation}
so $k_{\rm eff}+k'_{\rm eff}=k$ and
$\tau\simeq1/(2rk)$.  The exact solution is
\begin{equation}
\label{eq:3StateExSol}
  P_Y(t)=\frac{r}{2r+1} - \frac{1}{2} e^{-k\,t}+ \frac{1}{2 (2r+1)} e^{-(2 r+1)kt},
\end{equation}
whose linear‐in‐$t$ onset is given by
\begin{multline}
P_Y(t)=\tfrac12 k\Bigl(t-\underbrace{\tfrac1{(2r+1)k}}_{=\tau}\Bigr)+\mathcal{O}(t^2),\\
\text{for } (2r+1)^{-1}\ll kt\ll 1 \ ,
\end{multline}
fully consistent with $\tau\simeq1/(2rk)$.  
In other words, a large separation $r\!\gg\!1$ creates a clean intermediate-time
linear-growth window of slope $k/2$ with onset transient time $\tau\simeq1/(2rk)$.

\subsection{When Do We Need More States Instead?}

The number and structure of effective states depend on the interplay between the microscopic dynamical time scales and the time resolution of the observation. Hence, the minimal reduced Markov description is not uniquely determined \emph{a priori} by the microscopic definition of the model.
Depending on parameters and on the observation window, the same microscopic model may admit an effective two-state description, resolve a single slow intermediate mode (three-state), or display additional slow modes \emph{within} the intermediate region.
In the latter case, a single lumped state $I$ is no longer adequate, because the exit fluxes to $X$ and $Y$ are not determined by $P_I(t)$ alone.
A multi-state refinement is then required, for instance, when:
\begin{itemize}
  \item intra-$I$ mixing is not fast compared to the total escape out of $I$, so that the relative weights of the $I\to X$ and $I\to Y$ channels depend on the internal distribution within $I$;
  \item different regions (or microstates) within $I$ have different escape rates to $X$ and $Y$, producing multiple resolved time scales in $P_Y(t)$;
  \item the dynamics inside $I$ features additional long-lived traps, so that exits are governed by a sequence of metastable bottlenecks rather than a single quasi-equilibrated intermediate.
\end{itemize}
In these cases, splitting $I$ into substates $I_1,\dots,I_n$ is the minimal extension that retains a Markovian reduced description while capturing the newly resolved slow modes.

\subsection{Physical Aspects of the Three-State Model}

The three-state framework captures physical scenarios that are not faithfully represented by a direct two-state jump.
Different hierarchies among the rates $k_{X\to I}$, $k_{I\to X}$, $k_{I\to Y}$ and $k_{Y\to I}$ then correspond to distinct phenomenologies, reflected in the onset and crossover of $P_Y(t)$ and in the relative importance of recrossings, even within the same microscopic model.

\begin{itemize}
\item \textbf{Nucleation-and-growth in bistable systems.}
In many bistability scenarios, the transition proceeds with the rare creation of a \emph{precursor}---for instance, an incipient nucleus or a local rearrangement---which we collectively label as $I$.
The rate $k_{X\to I}$ sets the formation of such a precursor, while $k_{I\to Y}$ encodes its completion into the target phase.
Entering $I$ does not imply commitment: the precursor can either grow and complete the transition ($I\to Y$) or decay back to $X$, so that the crossover away from the quadratic short-time onset $P_Y(t)\sim t^2$ is controlled by the competition between these two outcomes.
A particularly simple limit is the \emph{fast cooperative completion} hierarchy $k_{I\to Y}\gg k_{X\to I}$ with $k_{I\to X}$ not dominant, in which the bottleneck is the formation of $I$ and trajectories typically commit rapidly once $I$ is reached.

\item \textbf{Fluctuation-driven reversals.}
If the backward channel $Y\to I$ is appreciable on the observation window, $Y$ is not effectively absorbing: returns to $I$ occur on time scales comparable to those over which $Y$ is first reached.
The dynamics then sustains a non-negligible probability flux from $Y$ back to $I$, producing repeated excursions $Y\to I\to Y$ rather than a one-way relaxation.
Within the present coarse-grained picture, this regime provides a minimal caricature of transitions in which external driving or environmental noise can repeatedly destabilise the nominal target state.

\item \textbf{Delayed Onset of $Y$ in Glassy or Spin-Glass Systems.}
In disordered landscapes, $I$ can represent a broad set of microstates within a larger metastable basin.
Once the system leaves $X$, part of the probability may follow comparatively fast routes through some substructures of $I$, while other substructures trap the dynamics for much longer times.
In practice, this heterogeneity can make the \emph{effective} progression toward $Y$ compete strongly with returns toward $X$, so that many excursions through $I$ are typically required before a successful commit occurs.
As a consequence, the growth of $P_Y(t)$ can display a pronounced slowing down already when $P_Y(t)\ll P_Y(\infty)$, without exhibiting a clean linear-growth window.
If the substates within $I$ have markedly different escape characteristics, a single lumped $I$ becomes insufficient and a more detailed multi-state description may be required.
Nevertheless, even the minimal three-state picture captures the idea that heterogeneous exploration of the intermediate region can substantially delay completion beyond a purely two-state (single-barrier) approximation.
\end{itemize}

Overall, the three-state model unifies these scenarios within a single kinetic template.
Varying the hierarchy among the rates continuously connects nucleation-limited regimes, rapid completion once $I$ is reached, and nonequilibrium cycling sustained by the $Y\to I$ channel.

For comparison with a setting where transition intermediates are structurally organized,
in Appendix~\ref{sec:app_2d_threestate} we present a three-state description of the $L=7$ 2D square-lattice Ising model at $\beta\!=\!1$.
Even in this scenario, the three-state description can still be quantitatively informative, supporting fixed-size surface tension estimates.


\section{The Zachary Karate Club Graph}
\label{sec:ZKC}

\begin{figure*}[ht!]
\includegraphics[]{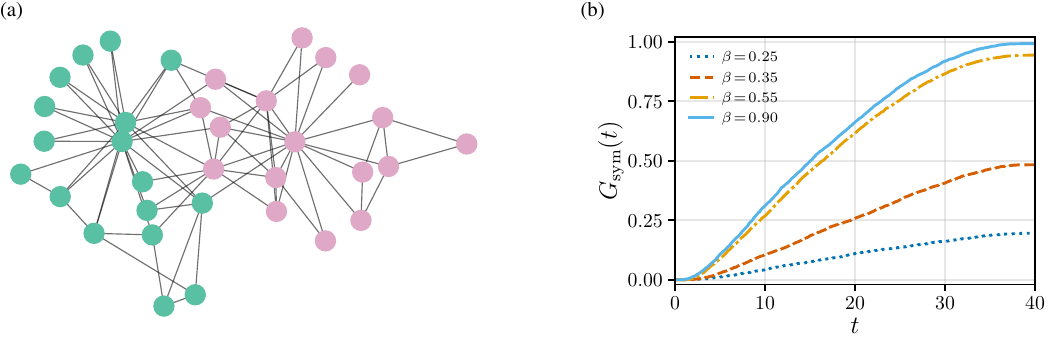}
    \vspace{-0.8cm}
\caption{
(a) Zachary Karate Club network with nodes colored according to a two-community partition. This defines the reference pattern $\underline{s}^{\rm comm}$ used in the overlap
$Q_{\rm comm}(t)=\sum_i s_i(t)\,s_i^{\rm comm}$.
(b) Ever-hit fraction
$G_{\rm sym}(t)=\mathbb P\bigl(\exists\,t'\le t:\,|Q_{\rm comm}(t')|\ge Q_\star\bigr)$,
i.e.\ the fraction of end-point conditioned trajectories that have visited the community-split sector at least once by time $t$.
We use $Q_\star=28$ (allowing at most $n_{\rm mis}=3$ mismatched spins out of $N=34$), and measure $G_{\rm sym}(t)$ at $T=40$ for the same representative $\beta$ values as in Fig.~\ref{fig:ZKC_regimes}.
As $\beta$ increases, $G_{\rm sym}(t)$ saturates to larger values, indicating that the community-split sector is visited by an increasingly large fraction of reactive trajectories.
}
\label{fig:ZKC_split_and_Gsym}
\end{figure*}

As an illustrative example, we consider the Zachary Karate Club (ZKC) network~\cite{Zachary1977}, which is a small heterogeneous social network with $N=34$ nodes and $E=78$ edges, often used as a benchmark example with a well-known modular structure \cite{GirvanNewman2002, DorogovtsevGoltsevMendes2008}; the network is shown in Fig.~\ref{fig:ZKC_split_and_Gsym}(a).
Despite its modest size, it displays a broad degree heterogeneity and a few bridging nodes between communities.
As we shall see, this heterogeneity can significantly affect the pathway and timing of the global spin flip.
We therefore use the ZKC as an illustrative real-world test for our framework before moving to random-graph ensembles.

We work in the ferromagnetic setup at zero external field, with $J_{ij}=1$ for each neighboring pair $(i,j)$ connected by an edge and $J_{ij}=0$ otherwise.
We study the $(-)\to(+)$ transition with hard boundary conditions $h_{in}\to -\infty$, $h_{out}\to +\infty$ and threshold $M^\star=20$, i.e., $m^\star \equiv M^\star/N\simeq 0.59$.
We consider trajectory lengths $T=10,\,15,\,20,\,25,\,30,\,35,\,40$ and perform TI up to $\beta_{\max}=0.90$.

Over the range of trajectory times $T$ analyzed here, $Z(t)$ displays distinct dynamical regimes as $\beta$ is varied.
These reflect the interplay between $\beta$-dependent residence times and the observational horizon $T$, leading to crossovers between the effective kinetic scenarios introduced above.
To quantify them, we fit $Z(t)$ within these scenarios, including a short-time shift $\tau_{\rm gran}$ accounting for a granular short-time scale before the reduced kinetic description applies. The construction mirrors the onset appearing in the quasi-static reduction of the three-state model, where the relaxation of the fast intermediate sector is absorbed into effective two-state rates. Here, $\tau_{\rm gran}$ collects the discreteness of single-spin updates, the finite time needed to complete a basin-to-basin crossing, and the relaxation between first entering an operational basin and reaching the corresponding effective kinetic state.

Quoted uncertainties are fit-only $1\sigma$ errors from a Jacobian-based covariance estimate; parameter correlations (notably in the saturation and three-state fits) are substantial and are not reported, so the errors should be read as indicative.
We identify four $\beta$ windows, illustrated in Fig.~\ref{fig:ZKC_regimes}:

\begin{figure*}[ht]
\includegraphics[]{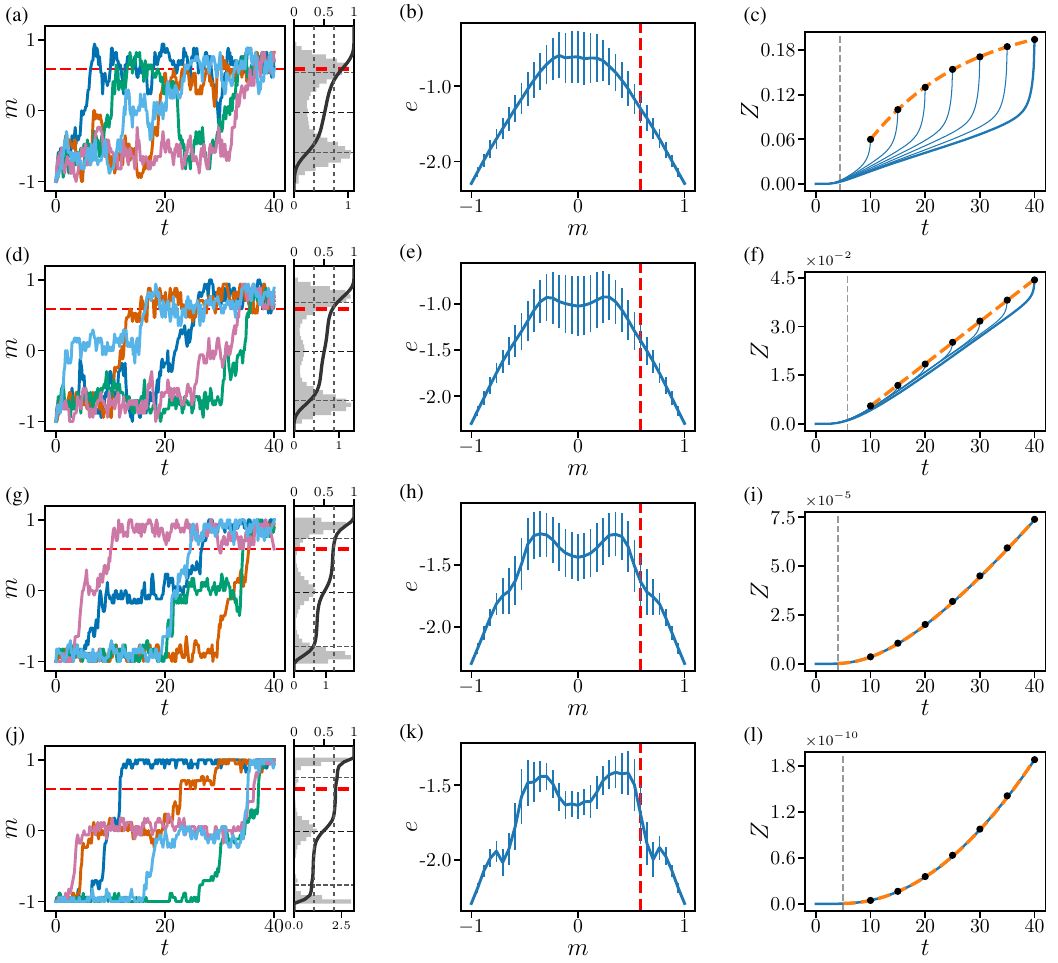}
\vspace{-0.65cm}
\caption{
Each row refers to one value of $\beta$ and displays (left) some magnetization trajectories $\{m(t)\}$, together with the density-normalized occupancy histogram (over all sampled trajectories) and its cumulative, showing the $(1/3,1/3,1/3)$ split for the high-$\beta$ regime discussed in Sec.~\ref{subsec:QuadraticGrowth}; (middle) the average energy and its dispersion at fixed magnetization; and (right) the thermodynamic--integration estimates at the sampled horizons $T$ (black markers), the corresponding reweighted family of curves $\{Z(t)\}_T$ obtained from Eq.~\eqref{eq:reweight} (blue solid curves), and the fits used in the analysis (dashed orange curves). The red dashed line in the left and center panels marks the threshold $m^\star$. For $\beta=0.25$ and $0.35$ (top two rows), the blue curves remain visibly distinct over their common time range, consistently with the persistence and basin-definition effects discussed in Sec.~\ref{ssec:rateTrick}; accordingly, the fit is performed \emph{only on the thermodynamic--integration points} (black markers). For $\beta=0.55$ and $0.90$ (bottom two rows), the blue curves are instead nearly perfectly superposed; in those cases, the fit is performed on the $Z(t)$ curve reconstructed from the single longest-$T$ TI run via the reweighting trick. From top to bottom, each row corresponds to:\\
{\scriptsize \textbullet} $\beta=0.25$ Fast saturation regime:
$Z_{\mathrm{fit}}(t)=c\,[1-\exp(-k_{\mathrm{sat}}\,(t-\tau_{\mathrm{gran}}))]$.\\
{\scriptsize \textbullet} $\beta=0.35$ Linear (two–state) regime:
$Z_{\mathrm{fit}}(t)=k\,(t-\tau_{\mathrm{fit}})$.\\
{\scriptsize \textbullet} $\beta=0.55$ Intermediate (three–state) regime:
$Z_{\mathrm{fit}}(t)$ given by Eq.~\eqref{eq:3StateExSol} with $t\to(t-\tau_{\mathrm{gran}})$.\\
{\scriptsize \textbullet} $\beta=0.90$ Quadratic onset:
$Z_{\mathrm{fit}}(t)=a\,(t-\tau_{\mathrm{gran}})^{2}$ for $t>\tau_{\mathrm{gran}}$.\\
An additive baseline $C$ is allowed in the intermediate and quadratic cases, but it is treated as a nuisance parameter: during optimization it is recomputed at every step to enforce zero-mean weighted residuals, while only $\tau_{\mathrm{gran}}$ (or $\tau_{\mathrm{fit}}$) and the kinetic constants ($k_{\mathrm{sat}}$, $k$, $r$, $a$) are varied by the solver. Best-fit values are reported in the text.
 }
    \label{fig:ZKC_regimes}
\end{figure*}

\begin{itemize}
    \item \textbf{\boldmath$\beta \lesssim 0.30$ — fast saturation.}  
    Here, $Z(t)$ bends rapidly towards a plateau (Fig.~\ref{fig:ZKC_regimes}(a)).
    We fit the TI estimates with a \emph{single-exponential saturation}
    \begin{equation}
    \label{eq:Zsatur}
      Z_{\text{fit}}(t)=
      c\Bigl[1-\exp\!\bigl(-k_{\mathrm{sat}}\,(t-\tau_{\mathrm{gran}})\bigr)\Bigr],
      \quad t>\tau_{\mathrm{gran}} .
    \end{equation}
    This behaviour is consistent with an effectively two-state dynamics in which the characteristic time $k_{\mathrm{sat}}^{-1}$ is not large compared to the observational window. In this regime, the single-TI reweighting procedure of Sec.~\ref{ssec:rateTrick} is not useful for rate extraction, because multiple switches are observed, and we rely on the TI points only.
    
    At $\beta=0.25$, we obtain as best-fit parameter values $k_{\mathrm{sat}}= 0.055\pm0.001$, $c= 0.226\pm0.002$, and $\tau_{\mathrm{gran}}= 4.45\pm 0.08$.
    The plateau value $c=\lim_{t\to\infty}Z(t)$ should be interpreted with care: it is the asymptotic probability of satisfying the \emph{operational} basin condition enforced by $(h_{\mathrm out},m^\star)$, i.e.\ of being \emph{above the threshold} $m^\star$, rather than the full equilibrium weight of the $(+)$ basin.
    In particular, if typical intra-basin fluctuations inside $(+)$ produce frequent dips below $m^\star$, then $c$ can substantially underestimate the actual basin weight $c_{(+)}$ even in a genuinely two-well, $\mathbb{Z}_2$-symmetric situation.
    
    More generally, the same saturating form of Eq.~\eqref{eq:Zsatur} can arise whenever all dynamical modes other than the slow $(-)\to(+)$ relaxation decay on time-scales $\lesssim\tau_{\mathrm{gran}}$: within our observational window one then sees a single effective saturation, while the precise value of the plateau depends on how sharply the basins are defined.
    This ambiguity is not a mere artifact of our fitting framework but reflects the intrinsic arbitrariness of state definitions and time-scale separation in finite (and/or non-mean-field) systems.

    As a diagnostic check of the threshold sensitivity, we repeated the low-$\beta$ analysis using a lower threshold $M^\star=6$ (Appendix~\ref{sec:app_ZKC_M6}).
    In zero field, the equilibrium distribution of $m$ is $\mathbb{Z}_2$-symmetric; in this high-temperature regime we can therefore interpret the plateau value as approximately
    $P_{\rm eq}(m\ge m^\star)=P_{\rm eq}(m\le -m^\star)=c_{M^\star=6}$, so that the probability mass outside the threshold-defined basins is
    $1-2c_{M^\star=6}$ (equivalently, each side contributes $0.5-c_{M^\star=6}$).
    We find $0.5-c_{M^\star=6}=0.036\pm0.004$,
    supporting an effectively two-state interpretation in this regime.

    \item \textbf{\boldmath$0.30 \lesssim \beta \lesssim 0.40$ — single jump.}  
      In this regime, $Z(t)$ grows linearly over the late-time window
      (Fig.~\ref{fig:ZKC_regimes}(d)), as expected for an effectively two-state process in an observation window where the fitted rate satisfies
      $kT\ll1$.  
      We use the \emph{linear form}
      \begin{equation}
        Z_{\text{fit}}(t)=k\,(t-\tau_{\mathrm{fit}})
      \end{equation}
      to fit the transition probabilities obtained from the TIs.
      In this regime the single time-shift $\tau_{\mathrm{fit}}$
      effectively captures both the granular transient and the two-state
      onset ($\tau_{\mathrm{fit}}\simeq\tau_{\mathrm{gran}}+\tau$
      in this regime); hence only $k$ and $\tau_{\mathrm{fit}}$ are fitted.

      The fitting procedure at $\beta=0.35$ yields $k= (1.30\pm 0.01)\times 10^{-3}$, $\tau_{\mathrm{fit}}= 5.8\pm 0.2$.
      Even though $kT\ll 1$, in this regime the single-TI reweighting trick is not directly applicable because post-entry fluctuations in the $(+)$ well yield non-negligible intervals with $m<m^\star$ (hence $\chi_{(+)}=0$), which spoils the smooth behavior of $\langle \chi_Y[\underline s(t)] \rangle$ near $t\simeq T$.
      A simple correction is possible (see Sec.~\ref{sec:RRG} and Appendix~\ref{sec:app_moderateBeta}), but here we instead rely on the multiple TI points available.

\item \textbf{\boldmath$0.40 \lesssim \beta \lesssim 0.65$ — intermediate state.}  
      In this regime, relevant trajectories pass through an intermediate state whose mean residence time is comparable to the observational time-scale $T$, and the mean residence times in the initial $(-)$ state and the final $(+)$ state are much larger than such a scale; therefore, 
      the full three–state kinetics is expected
      (Fig.~\ref{fig:ZKC_regimes}(g)). Once trajectories reach a magnetization $m>m^\star$, yielding $\chi_Y=1$, they do not leave that region of configuration space within our observation window, and the reweighting trick can be used reliably and straightforwardly.
      We fit the derived continuous curve with Eq.~\eqref{eq:3StateExSol}, replacing \(t\) by \(t-\tau_{\rm gran}\) as discussed above. We also allow for a small additive constant \(C\), treated as a nuisance baseline for residual coarse-graining mismatch and recomputed at fixed kinetic parameters by enforcing zero weighted-mean residuals; it is compatible with zero within the fit variability.
    The fit at $\beta=0.55$ yields $k= (7.18\pm0.005)\times 10^{-6}$, $r= (3.86\pm0.04)\times 10^{3}$ and $\tau_{\mathrm{gran}}= 3.9\pm0.1$.

\item \textbf{\boldmath$0.65 \lesssim \beta \lesssim 0.90$ — quadratic onset.}  
      Now the mean residence time in the intermediate state also exceeds the observation window: the early‐time quadratic regime dominates
      (Fig.~\ref{fig:ZKC_regimes}(j)).
          \clearpage \noindent
      We therefore fit  
      \begin{equation}
        Z_{\text{fit}}(t)=
           a\,(t-\tau_{\mathrm{gran}})^{2},
           \qquad t>\tau_{\mathrm{gran}} .
      \end{equation}
    Allowing again for the nuisance baseline \(C\) gives a value compatible with zero and does not affect the reported best-fit parameters, which at $\beta=0.90$ are $a\simeq (1.51\pm0.03)\times 10^{-13}$ and $\tau_{\mathrm{gran}}= 4.7\pm0.1$.

\end{itemize}

The modular structure of ZKC leaves a clear signature on the path ensembles.
In the $\beta$ range where a three-state description becomes necessary, the magnetization
histograms in Fig.~\ref{fig:ZKC_regimes} show an extended residence of the
dynamics around $m\simeq 0$; at $\beta=0.90$ the corresponding cumulative curve develops sharp steps at probabilities $\simeq 1/3$ and $\simeq 2/3$, in direct correspondence with the $(1/3,1/3,1/3)$ splitting discussed in Sec.~\ref{subsec:QuadraticGrowth}.
This phenomenology suggests that the activated flip is frequently mediated by a long-lived \emph{community-split sector}, in which the two main modules are magnetized in opposite directions, before the final relaxation into the $(+)$ well.

To quantify the prevalence and timing of this sector with an \emph{operational} marker---insensitive to microscopic rearrangements within each module---we fix a reference pattern
$\underline{s}^{\rm comm}\in\{\pm1\}^N$ encoding the ZKC split into two communities (Fig.~\ref{fig:ZKC_split_and_Gsym}(a); see also Refs.~\cite{GirvanNewman2002,DorogovtsevGoltsevMendes2008}).
For each $\beta$, we evaluate the diagnostic on the same end-point conditioned path ensemble used to measure $Z(t)$.
Given a trajectory $\underline{s}(t)$, we monitor its proximity to the community-split sector through the extensive overlap
\begin{equation}
  Q_{\rm comm}(t)\equiv \sum_{i=1}^N s_i(t)\,s_i^{\rm comm}.
\end{equation}
Large values of $|Q_{\rm comm}(t)|$ indicate that the two modules are magnetized in opposite directions, up to a small number of mismatches with respect to the reference split.
We declare that the trajectory has entered this sector when $|Q_{\rm comm}(t)|\ge Q_\star$, with $Q_\star=28$ (equivalently, allowing at most $n_{\rm mis}=3$ mismatched spins out of $N=34$).
We then consider the cumulative \emph{ever-hit} fraction
\begin{equation}
  G_{\rm sym}(t)\equiv \mathbb P\Bigl(\exists\,t'\le t:\ |Q_{\rm comm}(t')|\ge Q_\star\Bigr),
\end{equation}
i.e.\ the fraction of trajectories that have approached the community-split sector at least once by time $t$; by construction, this definition is insensitive to brief threshold crossings.

Fig.~\ref{fig:ZKC_split_and_Gsym}(b) reports $G_{\rm sym}(t)$ at $T=40$ for the same representative values of $\beta$ shown in Fig.~\ref{fig:ZKC_regimes}.
As $\beta$ increases, $G_{\rm sym}(t)$ rises earlier and saturates to larger values, approaching unity in the activated low-temperature cases $\beta=0.55$ and especially $\beta=0.90$.
Within the end-point conditioned path ensemble, this indicates that essentially all reactive trajectories visit the community-split sector at least once before completing the $(-)\!\to(+)$ event, providing direct support for the three-state interpretation suggested by the magnetisation histograms (Fig.~\ref{fig:ZKC_regimes}).
At $\beta=0.35$ and $\beta=0.25$, instead, $G_{\rm sym}(t)$ saturates
well below unity: under the strict $n_{\rm mis}\leq3$ definition, only a fraction of the conditioned trajectories are classified as ever visiting the community-split neighbourhood. This is expected in the fast-saturation and single-jump regimes discussed above, where the transition is less organized around a long-lived intermediate state. The nonzero value of $G_{\rm sym}$ nevertheless shows that the modular sector remains visible in a non-negligible fraction of conditioned trajectories.
As a sanity check, repeating the same construction with an uncorrelated reference pattern yields a signal compatible with zero over the same window (not shown), indicating that the effect is specifically tied to the ZKC modular partition.

On time scales where leaving the $(+)$ basin after first entry is negligible, $Z(t)$ can be interpreted as the cumulative probability that the transition has been completed by time $t$. Moreover, once the dynamics has rapidly lost memory within the $(-)$ basin (so that escape statistics become approximately stationary after a short transient), the derivative $dZ/dt$ provides direct access to the completion-time statistics (or, more generally, to its convolution with the escape-time statistics), and can be used as a diagnostic of intermediate-state kinetics; further discussion is reported in Appendix~\ref{ssec:app_transition_times}.

Overall, ZKC provides a compact illustration of how degree heterogeneity and modular structure generate intermediate-time kinetics beyond a two-state picture, and it delineates the $\beta$-regimes in which our procedure yields a reliable $Z(t)$ for quantitative analysis.

Representative trajectory animations for the ZKC runs are available as Supplementary Videos \cite{thisWork_Zenodo}.

\section{Random Regular Graphs}
\label{sec:RRG}

Random regular graphs (RRG) provide a convenient baseline to include disorder into a ferromagnetic problem.
The fixed degree $c$ removes degree heterogeneity, a major source of variability, while still leaving non-trivial sample-to-sample fluctuations in the graph topology.
We focus on $c=3$ for computational reasons.

\begin{figure*}[ht!]
\includegraphics[]{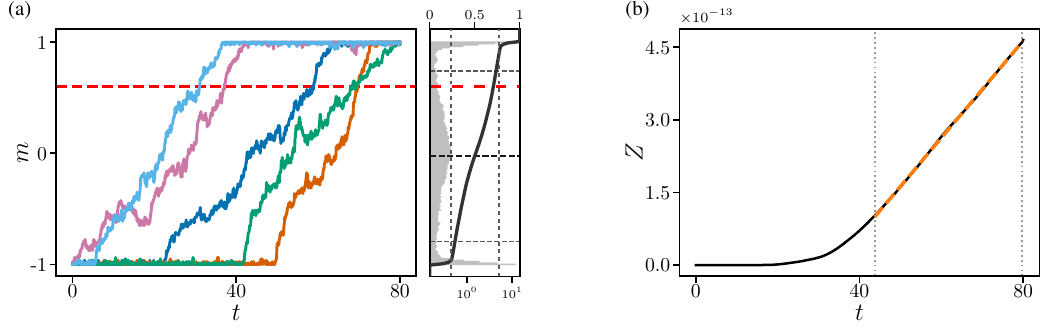}
 \vspace{-0.8cm}
    \caption{
    Instance of a random regular graph (RRG) with $c=3$, $N=160$, at $\beta=1$, with
    $h_{\mathrm{in}}=-\infty$, $h_{\mathrm{out}}=+\infty$ and $m^\star=0.6$ ($M^\star=96$), using total trajectory length $T=80$.
    \textbf{(a)} Representative magnetization trajectories $m(t)$ together with the
    density-normalized histogram of occurrences (log scale), computed over the
    \emph{full set of TPS-sampled trajectories used in the analysis} (not only the few
    paths shown). The red dashed horizontal line marks the threshold $m^\star$.
    Dashed horizontal lines (black/grey) mark the sample mean $\langle m\rangle$
    and $\langle m\rangle\pm\sigma_m$ of the histogram.
    Vertical guide lines on the histogram-density axis mark the reference cumulative levels $\frac12(1\mp\tau/T)$ expected under the crude estimate that a fraction $\tau/T$ of the trajectory time is spent in the jump, while the remaining time is equipartitioned between the two ferromagnetic basins.
    \textbf{(b)} Transition probability $Z(t)$ reconstructed from a single long-$T$ TI run via
    the reweighting identity of Sec.~\ref{ssec:rateTrick}; the dashed orange line is the late-time fit that identifies the linear-growth regime of $Z(t)$ and provides the operational estimates of the transition rate (slope) and of the onset/transient time $\tau$, here $\tau\simeq 43$.
    }
    \label{fig:mAndChiVsT_LT_RRG}
\end{figure*}
\begin{figure*}[ht!]
\includegraphics[]{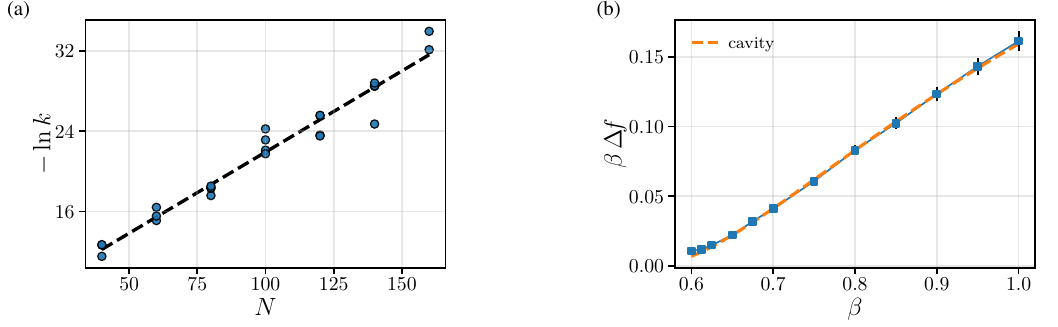}
\vspace{-0.75cm}
    \caption{
    Random regular graphs (RRGs) with connectivity $c\!=\!3$ at zero field, for the ferromagnetic Ising model with couplings $J_{ij}\!=\!1$ on edges (and $J_{ij}\!=\!0$ otherwise).
    \textbf{(a)} Transition rates $k_{-\to +}$ extracted at $\beta\!=\!1$ as a function of system size $N$ (one marker per graph instance), together with the linear fit in Eq.~\eqref{eq:fitRate} (dashed line).
    The slope yields the intensive free-entropy barrier $\beta \Delta f(\beta)$.
    By $\mathbb{Z}_2$ symmetry at zero field one has $k_{-\to +}\!=\!k_{+\to -}$.
    \textbf{(b)} Extracted values for $\beta \Delta f(\beta)$ over the investigated inverse-temperature range (markers) and the comparison with the corresponding \emph{thermodynamic} (or static) barrier computed via the cavity method (dashed line).
    }
    \label{fig:barrierExtractionAndBarrierVsBet}
\end{figure*}

We work in the ferromagnetic setting with couplings $J_{ij}=1$ on each graph edge $(i,j)$ and $J_{ij}=0$ otherwise.
We run the TPS algorithm to study $(-)\to(+)$ transitions on independent RRG instances, with sizes
$N\in\{40,60,80,100,120,140,160\}$.
We consider a total of 23 graphs; the number of graphs for each size is $\{3,
3,
4,
4,
4,
3,
2\}$.
We use hard boundary conditions $h_{\rm in}=-\infty$, $h_{\rm out}=+\infty$ and a basin threshold $m^\star=0.6$ (equivalently $M^\star=m^\star N$), and we collect data up to $\beta=1.0$.
For each graph instance $g$ we fix a single trajectory length $T$ by requiring that, at the lowest temperature considered ($\beta=1$), the reconstructed transition probability $Z(t)$ exhibits a clear linear-growth window.
The same $T$ is then used for all $\beta\le 1$ on that instance; in practice, this guarantees $T>\tau$ for the whole explored range, so that a linear-growth window can be identified whenever the dynamics is in the two-state/activated regime.
Transition rates are extracted from the slope of $Z(t)$ as explained in Sec.~\ref{ssec:samplingTrajectories}.
Representative thermodynamic-integration curves are reported in Appendix~\ref{ssec:app_TI_procedure}.

For the largest $\beta$ values (lowest temperatures), the two ferromagnetic basins are narrow enough that configurations with
$|m(\underline{s})|<m^\star=0.6$ are essentially visited only during the actual $(-)\!\to\!(+)$ jump,
see Fig.~\ref{fig:mAndChiVsT_LT_RRG}(a).
In this regime the threshold indicator $\chi_{(+)}(t)=\Theta(m(t)-m^\star)$ provides a robust operational signature of state occupancy,
and the transition probability $Z(t)$ displays the expected post-transient linear behavior,
Fig.~\ref{fig:mAndChiVsT_LT_RRG}(b).

At smaller $\beta$, instead, intra-basin fluctuations can induce repeated threshold recrossings even without a true basin-to-basin transition.
A rate extraction remains possible when the characteristic recrossing time $\tau_{\rm rec}$ is well separated from the mean transition time $k^{-1}$,
namely $\tau_\text{rec}\ll T \ll k^{-1}$ (i.e.\ $kT\ll1$), so that fast recrossings average out while the slow transition controls the long-time growth of $Z(t)$.
The operational consequences and a practical correction strategy are discussed in Appendix~\ref{sec:app_moderateBeta}.

Applying the above extraction across the $23$ RRG instances yields, for each investigated $\beta$, a set of rates $\{k_{-\to +}^{(g)}(\beta)\}_g$, where the index $g$ runs over the graphs.
We then test the activated size-scaling introduced in Eq.~\eqref{eq:ArrhLaw} by fitting the rates at a fixed $\beta$ according to the law
\begin{equation}
  \ln\left(k_{-\to +}(\beta,N)\right)=\ln A(\beta)-\beta\,N\,\Delta f(\beta)\,,
  \label{eq:fitRate}
\end{equation}
where $A(\beta)$ is a prefactor and $\Delta f(\beta)$ is the intensive free-energy barrier.
Fig.~\ref{fig:barrierExtractionAndBarrierVsBet}(a) shows the fit for $\beta=1$; its slope provides $\beta\,\Delta f(\beta)$,
while the intercept encodes $\ln A(\beta)$.
Repeating the extraction over the available $\beta$ values yields the barrier curve $\beta\,\Delta f(\beta)$ reported in
Fig.~\ref{fig:barrierExtractionAndBarrierVsBet}(b).

We compare the obtained dynamical estimates with a \emph{static} homogeneous Bethe (cavity) calculation for the ferromagnetic Ising model on RRGs, equivalently the homogeneous fixed-point branch of the corresponding Belief Propagation equations~\cite{bernaschi2021we,mezard_montanari}. The resulting Bethe free-energy density can be parametrized by the magnetization; the theoretical intensive barrier is identified as the free-energy difference between the symmetric point at $m=0$---the maximum along this homogeneous branch---and the ferromagnetic minimum at $m_{\rm FM}(\beta)$,
\begin{equation}
  \beta\,\Delta f_{\rm BP}(\beta)=\beta\,[f_{\rm Bethe}(0;\beta)-f_{\rm Bethe}(m_{\rm FM}(\beta);\beta)].
\end{equation}
The agreement between the TPS-based extraction and the static Bethe prediction (dashed curve in Fig.~\ref{fig:barrierExtractionAndBarrierVsBet}(b)) is very good across the explored range; the smallest-barrier regime is also the one in which threshold recrossings become relevant and subleading finite-$N$ effects are hardest to exclude (see Appendix~\ref{sec:app_moderateBeta}).

\section{Erdős–Rényi graphs}
\label{sec:ER}

\begin{figure*}[ht!]
\includegraphics{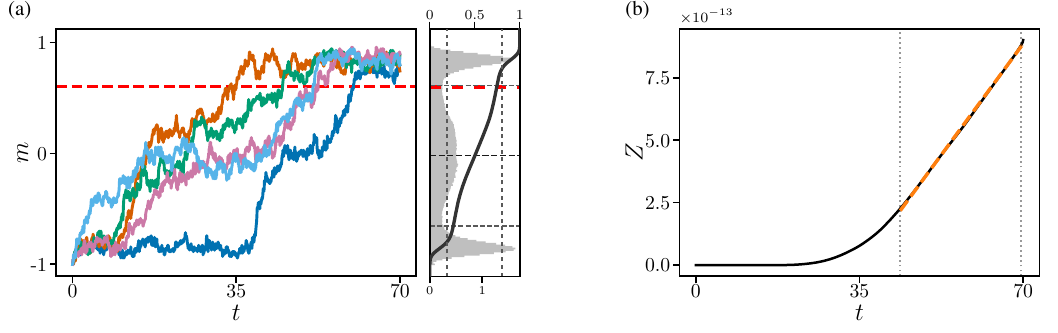}
    \vspace{-0.75cm}
    \caption{
    Instance of an Erd\H{o}s--R\'enyi (ER) graph with $c=3$, $N=140$, at $\beta=1$, with
    $h_{\mathrm{in}}=-\infty$, $h_{\mathrm{out}}=+\infty$ and $m^\star=0.6$ ($M^\star=84$), using total trajectory length $T=70$.
    \textbf{(a)} Representative magnetization trajectories $m(t)$ together with the density-normalized histogram of occurrences, computed over the TPS-sampled trajectory ensemble.
    The red dashed horizontal line marks the threshold $m^\star$.
    Vertical guide lines on the histogram axis indicate the reference cumulative levels $\frac12(1\mp\tau/T)$ expected under the crude estimate that a fraction $\tau/T$ of the trajectory time is spent in the jump, while the remaining time is equipartitioned between the two ferromagnetic basins.
    \textbf{(b)} Transition probability $Z(t)$ reconstructed from a single long-$T$ TI run via the reweighting identity of Sec.~\ref{ssec:rateTrick}; the dashed line highlights the linear-growth regime of $Z(t)$ and provides the operational estimates of the transition rate (slope) and of the onset/transient time $\tau$, here $\tau\simeq 41$.
    }
    \label{fig:chiAndMVsT_LlT_ER}
\end{figure*}

Having established the method on random regular graphs, we now turn to a more heterogeneous setting in which the local connectivity is not fixed. We consider Erd\H{o}s--R\'enyi (ER) random graphs, where each of the $\binom{N}{2}$ possible edges is present independently with probability $p$, so that vertex degrees fluctuate from site to site. To obtain a sparse ensemble with size-independent average connectivity, we set $p=c/(N-1)$; in the thermodynamic limit, the degrees become i.i.d.\ Poisson random variables with mean $c=p(N-1)$. This structural difference has concrete consequences already at the level of static finite-size corrections: for Ising models on ER graphs, connectivity fluctuations generate additional contributions beyond the loop-controlled corrections that dominate on random regular graphs (e.g.\ in the form of short open-chain terms in a $1/N$ expansion)~\cite{FRT2014}.
In the following, we focus on $c=3$ (for computational convenience) and study the ferromagnetic Ising model with couplings $J_{ij}=1$ on graph edges and $J_{ij}=0$ otherwise.

We run the TPS algorithm to study $(-)\to(+)$ transitions on independent ER instances, with sizes
$N\in\{40,60,80,100,120,140,160,180,200\}$.
We consider a total of 50 graphs; the number of graphs for each size is 
$\{
6,
6,
6,
7,
7,
6,
5,
4,
3
\}$.
We use hard boundary conditions $h_{\rm in}=-\infty$, $h_{\rm out}=+\infty$ and a basin threshold $m^\star=0.6$ (equivalently $M^\star=m^\star N$).
For each graph we choose the total trajectory length $T$ large enough that $Z(t)$ reconstructed at the reference inverse temperature $\beta=1$ displays a clear post-transient linear-growth window; this allows us to extract the transition rate from its slope as explained in Sec.~\ref{ssec:samplingTrajectories} (and we operationally enforce the condition $T>\tau(\beta=1)$ a posteriori, as discussed below).
Thermodynamic integration is performed on an instance-dependent inverse-temperature grid, always including $\beta=1$ and, when needed, extending to larger $\beta$ values, as discussed below.
Representative thermodynamic-integration curves are reported in Appendix~\ref{ssec:app_TI_procedure}.

From the single-instance perspective, ER transition trajectories retain the same qualitative structure as in the RRG case, as shown in Fig.~\ref{fig:chiAndMVsT_LlT_ER}(a): long stretches of intra-basin relaxation and fluctuations are interrupted by a rare excursion that brings the system to the opposite basin. A distinctive feature of ER instances is that the heterogeneous local environment (including weakly connected and isolated vertices) prevents the magnetization from strictly saturating to $|m(\underline{s})|=1$, so that sizable intra-state fluctuations of $m$ may persist even at large $\beta$. We simulate the full ER graph, i.e.\ we do not restrict the dynamics to a connected component. The extraction of transition rates $k$ from the late-time behavior of $Z(t)$, shown in Fig.~\ref{fig:chiAndMVsT_LlT_ER}(b), follows the same protocol as for RRGs; in the moderate-$\beta$ regime, where intra-state fluctuations in the final basin may yield $m<m^\star$, we apply the prescriptions detailed in Appendix~\ref{sec:app_moderateBeta}.

\begin{figure*}[ht!]
\centering
\includegraphics[]{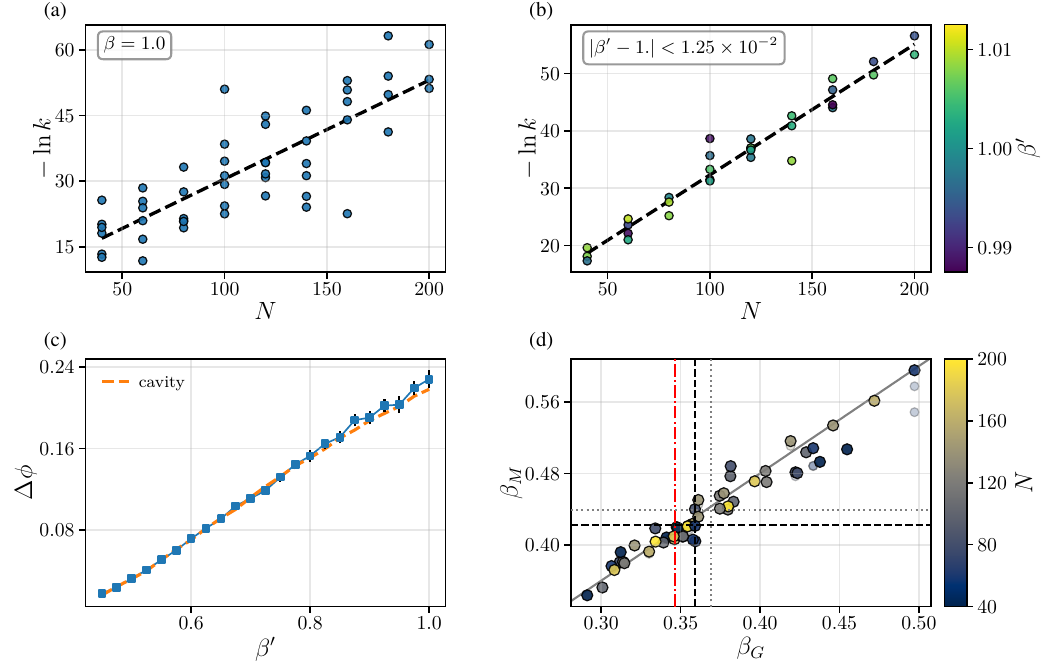}
    \vspace{-0.55cm}
    \caption{
Results for Erd\H{o}s--R\'enyi graphs with average connectivity $c=3$ (couplings $J_{ij}=1$ on edges, $J_{ij}=0$ otherwise).
\textbf{(a)} Attempt to extract the free-entropy barrier $\Delta\phi(\beta)=\beta\Delta f(\beta)$ from unscaled transition rates at the common inverse temperature $\beta=1$. Graph-to-graph fluctuations obscure the size dependence.
\textbf{(b)} After rescaling each graph’s inverse temperature by its intrinsic scale $\beta_g=\beta_M(g)$, data points assigned to the bin $\beta'=1$ collapse onto an approximately linear behavior; the slope provides an estimate of $\Delta\phi(\beta'=1)$. Colors encode the rescaled values $\beta'=\beta\,\overline{\beta_M}/\beta_M(g)$ of the individual data points.
\textbf{(c)} Free-entropy barrier $\Delta\phi(\beta')=\beta'\Delta f(\beta')$ as a function of the rescaled inverse temperature $\beta'$. All graph sizes are used. The dashed orange line is the Bethe (cavity) prediction for the ER ensemble, computed by population dynamics as described in the main text.
\textbf{(d)} Validation of the rescaling proxies: linear correlation between $\beta_M$ from thermodynamic integration and the graph-theoretic estimate $\beta_G$ defined in Eq.~\eqref{eq:betaG_def}. The solid line shows the proportionality law $\beta_M = 1.20\,\beta_G$. Opaque points correspond to retained TI datasets ($T>\tau(\beta=1)$); pale points (excluded) are exploratory datasets with $T\le\tau(\beta=1)$. Colors encode system size $N$. Horizontal and vertical dotted gray lines report the means of $\beta_M$ and $\beta_G$, respectively, while dashed black lines show the corresponding sample medians. The mean-median mismatch is driven by a small subset of instances, predominantly at the smallest sizes, consistent with enhanced finite-size variability. The red dot-dashed vertical line marks the $N\to\infty$ estimate $\beta_G^{(\infty)}=\operatorname{arctanh}(1/3)\simeq0.3466$ (for $J=1$).
}
    \label{fig:er_rateExtraction}
\end{figure*}

In the ER case, however, attempting to extract a free-entropy barrier from fits of $-\ln k$ versus $N$ at fixed $\beta$ does not reveal a clear linear trend across our graph instances. As shown in Fig.~\ref{fig:er_rateExtraction}(a), graph-to-graph fluctuations at a given size look very severe, making the estimate of the slope highly uncertain.
An analogous graph-to-graph spread already emerges upon inspection of the thermodynamic-integration curves
(Appendix~\ref{ssec:app_TI_procedure}). For each instance, the transition probability measured at fixed total trajectory length $T$ displays a pronounced maximum as a function of $\beta$, and the position of this maximum varies substantially across graphs (unlike in the RRG case). We denote by $\beta_M(g)$ the inverse temperature at which this maximum occurs for graph $g$; since $U=\partial_\beta \ln Z$ in our TI scheme, see Eq.~\eqref{eq:U_def}, $\beta_M$ equivalently corresponds to the point where $\langle \mathcal U[\underline{s}(t)]\rangle=0$.

This instance dependence provides an operational way to interpret the failure of the Arrhenius analysis at a fixed $\beta$. In that analysis, the slope is extracted from rates measured on different ER instances at the same bare inverse temperature.
The drift of $\beta_M(g)$ indicates instead that each instance carries its own characteristic inverse-temperature scale, so that the same bare $\beta$ need not define an equivalent comparison point across graphs.
We therefore introduce a graph-dependent scale $\beta_g$ and test whether the measured rates are better organized when compared at fixed reduced inverse temperature $\beta/\beta_g$.
This interpretation is consistent with the expected exponential dependence of activated rates on an extensive, temperature-dependent cost.
If $\beta_g$ captures the leading instance dependence of the temperature scale governing this cost, then comparing graphs at fixed $\beta/\beta_g$ places them at comparable reduced temperatures and should therefore reduce differences in the exponential contributions to the rates.
At fixed bare $\beta$, even modest differences in $\beta_g$ may instead contribute to the large spread in $\ln k$ observed in Fig.~\ref{fig:er_rateExtraction}(a). The adequacy of the rescaling is assessed a posteriori by the quality of the data collapse across the range of rescaled inverse temperatures $\beta'$.

Since the target event is the formation of a magnetized basin, a meaningful proxy should probe the ferromagnetic ordering scale of the graph. We consider:
\begin{itemize}
    \item the inverse temperature $\beta_M(g)$ obtained from thermodynamic integration as the position of the maximum of the transition probability (equivalently through the condition $\langle \mathcal U[\underline{s}(t)]\rangle=0$);
    \item a static estimate $\beta_G(g)$, obtained from the linear stability of the paramagnetic BP/cavity fixed point on the specific graph~\cite{mezard_montanari}. At zero external field this fixed point has vanishing cavity messages. Linearizing the BP update around it, the evolution of small perturbations is governed by the non-backtracking matrix $B_g$~\cite{ref_NB_BP_stability_1,ref_NB_BP_stability_2}, with a factor $\tanh(\beta J)$ associated with each non-backtracking step; the instability condition therefore reads
  \begin{equation}
    \tanh(\beta_G J)\,\lambda_{\max}(B_g)=1,
  \end{equation}
  hence for $J=1$:
  \begin{equation}
    \beta_G(g)=\operatorname{arctanh}\!\left(\frac{1}{\lambda_{\max}(B_g)}\right).
    \label{eq:betaG_def}
  \end{equation}
\end{itemize}

The appeal of $\beta_M$ is operational: it is extracted directly from the TPS/TI workflow and does not require any additional graph-dependent input beyond the transition data. It also has a direct physical interpretation. At fixed horizon $T$, the transition probability is suppressed on both sides of its maximum: at small $\beta$ the magnetized target sector $m\ge m^\star$ has a small equilibrium weight, whereas at large $\beta$ the crossing becomes activated and rare on the simulated time scale. The maximum therefore marks the crossover between these two regimes and provides a dynamical estimate of the graph-dependent ordering scale.
The appeal of $\beta_G$ is instead that it is independent of the TPS construction, and depends only on the graph topology via the non-backtracking matrix; yet, its definition is rooted in an $N\to\infty$ stability criterion, and finite-size effects may be non-trivial on individual instances.

Since $\beta_M(g)$ is defined at fixed trajectory length $T$---and, in principle, also depends on the parameters setting the stochastic measure, $h_{\rm in}$, $h_{\rm out}$ and $m^\star$---we need an operational criterion to ensure that it is extracted in a regime where finite-$T$ effects are under control. Empirically, we observe that, once the trajectory length satisfies $T>\tau(\beta=1)$, further increases of $T$ have only a mild impact on the inferred $\beta_M$ (over the sizes and parameters explored here). Here $\tau(\beta=1)$ is the onset time of the linear-growth regime of $Z(t)$ reconstructed at our reference inverse temperature $\beta=1$ (always included in our TI grids) via the reweighting identity of Sec.~\ref{ssec:rateTrick}. We then retain for the analysis only those TI datasets satisfying $T>\tau(\beta=1)$; shorter exploratory datasets ($T\le\tau(\beta=1)$) are excluded.

Fig.~\ref{fig:er_rateExtraction}(d) then puts the two candidates on equal footing by plotting, for each ER instance, the pair $(\beta_G,\beta_M)$ (markers colour-coded by the system size $N$). For all investigated sizes ($40\le N\le 200$) the retained datasets cluster around an approximately linear relation,
\begin{equation}
  \beta_M \;\simeq\; 1.20\,\beta_G,
  \label{eq:betaCorr}
\end{equation}
with a relative scatter below $\sim 5\%$ once $N\gtrsim 100$. The few outliers lie systematically below the main trend and correspond to the excluded exploratory datasets ($T\le \tau(\beta=1)$), shown as pale markers.

Given the tight correlation in Eq.~\eqref{eq:betaCorr}, rescaling $\beta$ by either $\beta_M(g)$ or $\beta_G(g)$ yields very similar instance-to-instance realignment of temperature scales. Conversely, because $\beta_M$ is extracted from the finite-$T$ transition dynamics, it may partly absorb subleading finite-size/kinetic effects beyond a pure temperature-scale variability: this tends to improve the collapse of rate curves over the accessible sizes, while the slope-level agreement with the static prediction can be slightly tighter when using a topology-only rescaling. We therefore adopt
\begin{equation}
  \beta_g \equiv \beta_M(g)
\end{equation}
as the instance-dependent rescaling and show in Fig.~\ref{fig:er_rateExtraction}(b) the rescaled data, which indeed collapse much better.
We report in Appendix~\ref{sec:app_betaG_rescaling} the topology-based rescaling using $\beta_G$ as a complementary check.

More precisely, we proceed as follows:
\begin{enumerate}
    \item For each graph $g$ we estimate $\beta_M(g)$ from the retained TI dataset (satisfying $T>\tau(\beta=1)$) selected according to the criterion described above, and set $\overline{\beta_M}\equiv \mathrm{median}_{g}\,\{\beta_M(g)\}$ (each graph counted once), a robust reference scale against uneven size sampling and occasional outliers.
  \item To compare data across different graphs we introduce the rescaled inverse temperature
  \begin{equation}
    \beta' \equiv \beta \,\frac{\overline{\beta_M}}{\beta_M(g)} .
    \label{eq:newBetas}
  \end{equation}
  \item The mapping in Eq.~\eqref{eq:newBetas} transforms the common discrete $\beta$-ladder used in the simulations to a graph-dependent set of $\beta'$ values, so that different instances typically do not share the same $\beta'$. We therefore introduce a uniform grid of inverse temperature $\{\tilde\beta_n=n\epsilon\}_{n\in\mathbb{N}_0}$ and bin all measurements with $\beta'\in[\tilde\beta_n-\epsilon/2,\tilde\beta_n+\epsilon/2]$, assigning them the representative value ${\beta'=\tilde\beta_n}$.
\end{enumerate}

In the following, we set $\epsilon=0.025$. To keep, as far as possible, the same set of instances represented across bins up to $\beta'=1$, we extend the TI grids to $\beta>1$ for graphs with $\beta_M(g)>\overline{\beta_M}$, so that the rescaled coverage remains comparable after applying Eq.~\eqref{eq:newBetas}.  

Fig.~\ref{fig:er_rateExtraction}(b) shows $-\ln k$ versus $N$ at the rescaled temperature $\beta'=1$: after the instance-dependent rescaling described above, the cloud of points collapses onto an approximately linear behaviour, whose fitted slope provides an estimate of the free-entropy barrier $\Delta\phi(\beta')=\beta'\,\Delta f(\beta')$ at $\beta'=1$.
Repeating the same binning-and-fit procedure for each rescaled $\beta'$ yields the barrier curve $\Delta\phi(\beta')$ shown in Fig.~\ref{fig:er_rateExtraction}(c). Because the rescaling in Eq.~\eqref{eq:newBetas} and the subsequent binning introduce an additional (instance-dependent) uncertainty on the effective $\beta$ values, a full propagation of errors would require accounting for both statistical fluctuations and rescaling-induced systematics; we do not attempt such a detailed analysis here. 

As a static benchmark, we compute the Bethe (cavity) prediction for the ER ensemble by population dynamics for a Poisson degree distribution with mean $c$~\cite{mezard_montanari,ref_PD_BP_2} (dashed curve in Fig.~\ref{fig:er_rateExtraction}(c)). In this setting, the population dynamics solves for the distribution of cavity fields entering the BP fixed point. We follow the ferromagnetic solution and evaluate its Bethe free entropy $\phi_{\rm FM}(\beta)$; the zero-message paramagnetic BP solution continued beyond its linear-stability threshold provides the symmetric saddle, with Bethe free entropy $\phi_{\rm PM}(\beta)$. The corresponding static free-entropy barrier reads
\begin{equation}
\Delta\phi_{\rm Bethe}(\beta)=\phi_{\rm FM}(\beta)-\phi_{\rm PM}(\beta).
\end{equation}

\section{Conclusions}

We studied activated transitions in the ferromagnetic Ising model on sparse, disordered graphs, which represent the simplest possible example of a disordered system with multiple metastable states.
Building on the Transition Path Sampling (TPS) and thermodynamic-integration framework of Ref.~\cite{ZMW_2012}, we extracted finite-size rates and barriers in regimes where transition probabilities are exponentially small in $N$, as in Eq.~\eqref{eq:ArrhLaw}. We reconstructed constrained transition probabilities and, whenever a late-time linear-growth window is present in the transition-probability curve, extracted effective transition rates from dynamical data. To make this rate-extraction procedure transparent in metastable regimes, we introduced an effective three-state kinetic picture $X\!\to\!I\!\to\!Y$ that isolates intermediate trapping through residence times and clarifies how a linear-growth regime can emerge after transient effects. We illustrated the resulting diagnostics on the heterogeneous Zachary Karate Club network, where the comparison between direct thermodynamic-integration points and time-reweighted reconstructions provides a concrete check of when the single-$T$ reweighting procedure is reliable. Turning to random regular graph (RRG) ensembles, the same pipeline yields consistent barriers and enables a direct comparison with the corresponding static Bethe (cavity) prediction. For Erd\H{o}s--R\'enyi graphs, instead, strong instance-to-instance variability at fixed $(N,\beta)$ obstructs a direct barrier extraction from unscaled rates; motivated by the empirical evidence that each instance carries an intrinsic temperature scale, we introduced an instance-dependent inverse-temperature rescaling that realigns transition data across graphs and enables an ensemble-level barrier extraction and comparison with the Bethe prediction. A robustness check based on a topology-only scale supports that these conclusions do not hinge on the specific operational proxy used to set the rescaling.

All rate extractions rely on finite observation windows and on an operational definition of initial and final basins. The time-reweighting reconstruction is powerful but must be applied only within regimes where its underlying persistence assumptions are validated by diagnostics; at moderate temperatures, post-entry fluctuations can violate strict persistence and require explicit corrections or alternative persistence indicators. Moreover, while the barrier extraction is organized around the large-$N$ scaling in Eq.~\eqref{eq:ArrhLaw}, accessible sizes may retain subleading finite-size and kinetic contributions that are difficult to disentangle from genuine disorder effects. In the presence of quenched heterogeneity, a further conceptual issue is that reactive trajectories may organize into competing mesoscopic ``channels'' that are not related by simple symmetries, so that establishing exhaustive exploration of the full trajectory measure can be intrinsically difficult at finite computational cost. This motivates the use of internal consistency checks and trajectory-level diagnostics alongside rare-event sampling, while recognizing that such diagnostics may still fail to detect particularly slow or highly non-local obstructions to mixing in trajectory space.

In addition to the checks explicitly reported in the main text, we performed internal robustness tests across independent TPS runs. In particular, we monitored the autocorrelation of the TI integrand $U$ along the path-MC chain as a function of Monte Carlo sweeps, and we verified compatible coarse trajectory statistics such as the magnetization histogram along paths (a proxy for state occupancy). These practical diagnostics support stability with respect to initialization and run-to-run fluctuations at the level relevant for rate extraction; at the same time, they do not rule out especially slow or highly non-local trapping mechanisms in trajectory space, for which identifying sharp trajectory observables \emph{a priori} is itself non-trivial.

Within these constraints, our results show that TPS combined with thermodynamic integration provides a viable and flexible route to activated-barrier extraction on sparse random topologies. More broadly, they highlight that quenched topological disorder can strongly affect the dynamical observables used to estimate rates, so that instance-aware strategies may be required even when the underlying thermodynamic description (and corresponding static predictions) is self-averaging in the thermodynamic limit. Finally, minimal coarse-grained kinetic pictures can play a useful role in separating transient-time effects from genuine changes in activated kinetics, thereby clarifying when a single-rate description is operationally meaningful.

Further extensions include developing systematic multi-$T$ thermodynamic-integration datasets and diagnostic criteria to control, and if necessary correct, the breakdown of the single-$T$ time-reweighting reconstruction in regimes with recrossings or weak persistence. It would also be valuable to incorporate a more explicit propagation of statistical and systematic uncertainties in the rescaled, ensemble-level barrier extraction for Erd\H{o}s--R\'enyi graphs, and to assess the robustness of the instance-dependent scaling strategy across different disorder classes and connectivities. Finally, the sampled reactive trajectories naturally provide access to genuinely path-dependent observables beyond rates and barriers (e.g.\ coarse measures of activity, state-occupancy functionals, or topological ``footprints'' of the crossing on the graph), which could be leveraged both to characterise channel heterogeneity and to design sharper diagnostics of possible trapping in trajectory space; exploring these directions, potentially in combination with enhanced trajectory-space moves (non-local updates, replica-exchange and tempering schemes), is an interesting avenue for future work.

\begin{acknowledgments}
We thank {Ada Altieri}, {Louise Budzynski}, {Patrick Charbonneau}, {Leticia F. Cugliandolo}, {Matteo Negri}, {Saverio Palazzi}, {Giorgio Parisi} and {Tommaso Rizzo} for many useful discussions related to this project.

This work has been supported by the “National Centre for HPC, Big Data and Quantum Computing”, Project CN\_00000013, CUP B83C22002940006, NRRP Mission 4 Component 2 Investment 1.4, Funded by the European Union - NextGenerationEU.
\end{acknowledgments}

AI-assisted tools were used by R.C. for language editing and checks of clarity.

\bibliographystyle{apsrev4-2}
\bibliography{biblio}

\appendix

\section{Computational advantage over direct dynamics}
\label{sec:app_compAdv}

We provide here an order-of-magnitude comparison illustrating the computational advantage of the TPS+TI strategy over \emph{plain} direct dynamical estimates for the rare transition rates targeted in this work.
As a representative benchmark, we focus on the ER instance corresponding to the largest value of $-\ln k$ in Fig.~\ref{fig:er_rateExtraction}(a) at $\beta=1$, for which the transition probability at the simulation time horizon $T=80$ is $Z(T)\simeq 1.5\times10^{-26}$.
Fitting the post-transient linear-growth window of $Z(t)$ yields the rate estimate $k\simeq 3 \times 10^{-28}$.

\paragraph{TPS+TI cost.}
For this instance, the total CPU time effectively used to produce the TPS+TI data entering the plots amounts to $\simeq 4$ months on a \texttt{Intel Core i7-12700K} CPU (reported as single-core time, i.e.\ as the sum of the wall clock times of the runs effectively used in the plots).
The same TI procedure yields the full transition-probability curve $Z(T;\beta)$ across the sampled $\beta$ values and, whenever a linear-growth window is resolved, the corresponding rate estimates.
In practice, statistics for the time-resolved observable $\chi$ entering the reweighting (the estimator introduced in Sec.~\ref{ssec:rateTrick} to reconstruct $Z(t)$ from a single long-$T$ run) are the dominant cost, while the scalar TI estimator $U=\partial_\beta\ln Z$ converges faster.
Standard Monte Carlo runs used for the high-temperature anchor/matching are not included in this estimate, as their cost is negligible compared to the TPS+TI production runs in the rare-event regime.

\paragraph{Direct event-driven baseline.}
As a baseline for \emph{direct} estimates, we consider the same continuous-time single-spin Glauber dynamics simulated with a standard event-driven scheme (Gillespie), i.e.\ straightforward direct dynamics without dedicated rare-event acceleration.
We run $n_0=10^5$ independent direct trajectories started in the $(-)$ basin and evolved up to the same time horizon $T=80$.
The corresponding wall clock time is $\simeq 44.8\,\mathrm{s}$, yielding an average cost per trajectory $t_{\rm traj}\simeq 4.48 \times 10^{-4}\,\mathrm{s}$ on the same hardware.
In these $n_0=10^5$ direct runs, we observe no completed transitions by $T=80$, consistently with the extreme smallness of $Z(T)$ inferred from TPS.

To estimate the transition rate $k$ directly, one may run $n$ independent event-driven trajectories up to a fixed observation time $T$ and measure, at each time $t$, the empirical transition probability $\widehat Z(t)=S(t)/n$, where $S(t)$ counts trajectories with $\chi_{(+)}(t)=1$ as defined in the main text (i.e.\ $m(t)\ge m^\star$).
A rate estimate would then be obtained by fitting the late-time linear-growth window of $\widehat Z(t)$ over a time interval of width $\Delta t_{\rm eff}$, chosen here as the width of the window identified in the TPS-based reconstruction of $Z(t)$ for this same instance at $\beta=1$, $\Delta t_{\rm eff}=40$.
At the level of order-of-magnitude counting, obtaining a meaningful slope estimate amounts to requiring a sufficient number of statistically resolvable transition events within that window; we quantify this by a target relative uncertainty $\delta$ on $k$.

In the rare-event regime $Z(t)\ll 1$, $S(t)$ is well-approximated as a Poisson variable with mean $nZ(t)$, so that, within the linear-growth regime, the expected increment over the fit window scales as
\begin{equation}
  \mathbb{E}\big[S(t_2)-S(t_1)\big]\;\simeq\; n\,k\,(t_2-t_1)\equiv n\,k\,\Delta t_{\rm eff}.
\end{equation}
Accordingly, the relative statistical uncertainty on the fitted slope scales as
\begin{equation}
  \frac{\delta k}{k}\;\sim\;\frac{1}{\sqrt{n\,k\,\Delta t_{\rm eff}}}\,,
\end{equation}
and achieving $\delta k/k=\delta$ requires $n_\delta \simeq 1/(\delta^2\,k\,\Delta t_{\rm eff})$.
In this approximation, $\delta$ corresponds to requiring an expected number of events $n\,k\,\Delta t_{\rm eff}\sim 1/\delta^2$ within the fit window.
Using the measured direct-dynamics cost per trajectory $t_{\rm traj}$ from the $n_0=10^5$ benchmark above, the corresponding wall time scales as $t_{\rm dir}\sim n_\delta\,t_{\rm traj}$.
This estimate is already optimistic for direct dynamics, since it assumes that the relevant linear-growth window is known in advance and does not include the additional sampling effort needed to identify it reliably from noisy direct-dynamics data.
For the present instance with $k\simeq 3 \times 10^{-28}$, taking $\Delta t_{\rm eff}=40$ as the width of the linear-growth window identified in the TPS-based reconstruction of $Z(t)$ at $\beta=1$, and $\delta=0.15$, yields $t_{\rm dir}\gtrsim 5\times 10^{16}$ years on a single core.

Altogether, the comparison shows that, for the rare rates probed here, the TPS+TI workflow makes feasible transition-rate estimates that would be effectively out of reach for direct dynamics at the system sizes considered.

\section{Thermodynamic integration}
\label{sec:app_TI}

\subsection{TI procedure and representative instances}
\label{ssec:app_TI_procedure}

\begin{figure*}[ht!]
\includegraphics[]{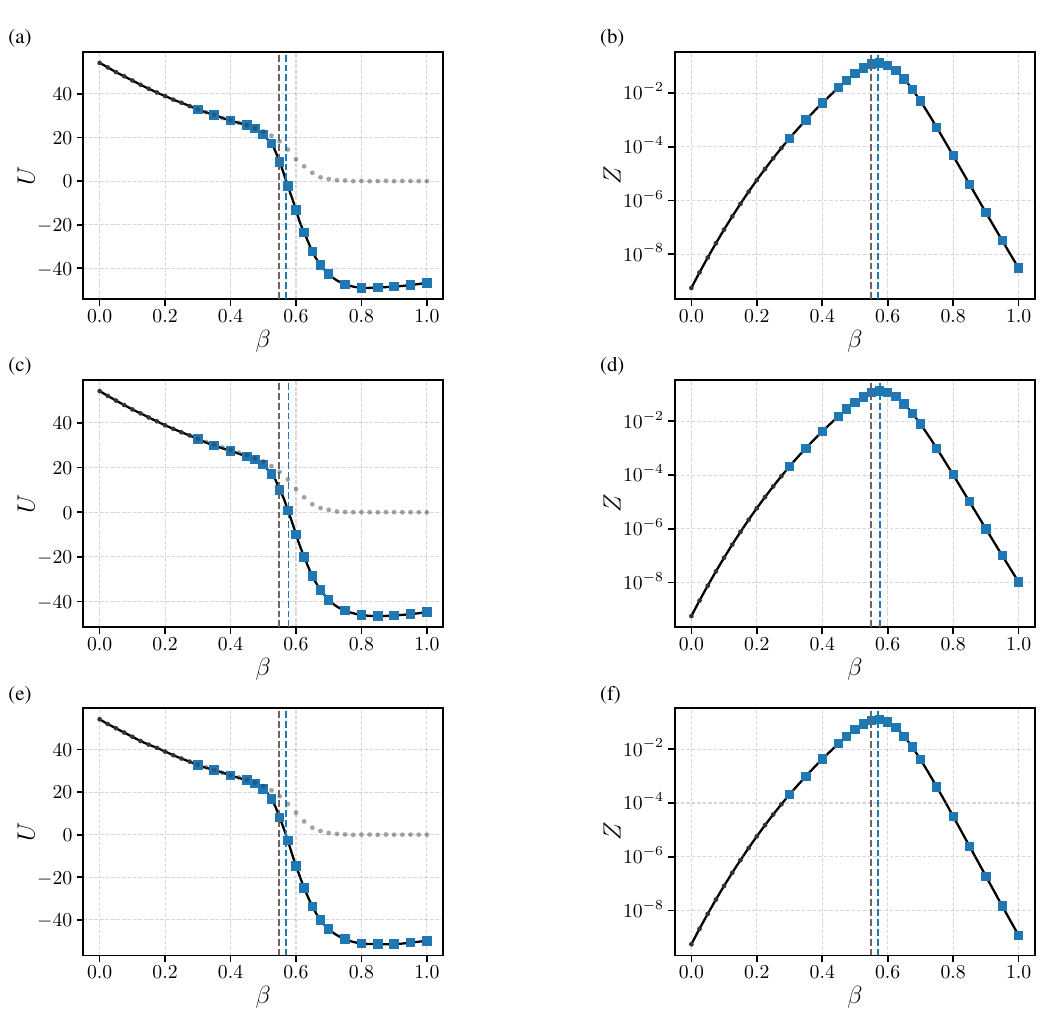}
\vspace{-0.5cm}
    \caption{
    Representative thermodynamic-integration datasets for random regular graphs (RRG, $c=3$) at fixed size $N=100$ and trajectory length $T=60$.
    Panels \textbf{(a)}, \textbf{(c)}, and \textbf{(e)} show the TI integrand $U(T;\beta)=\partial_\beta \ln Z(T;\beta)$ for three different quenched instances; TPS estimates are shown as squares, while standard-MC estimates are shown as circles.
    Opaque circles denote MC points actually used in the matched integrand, whereas faded circles indicate MC measurements beyond the matching point, shown for reference only.
    Vertical dashed lines mark the graph-dependent scales $\beta_G$ (grey) and $\beta_M$ (blue).
    Panels \textbf{(b)}, \textbf{(d)}, and \textbf{(f)} show the corresponding $Z(T;\beta)$, reconstructed through thermodynamic integration of the matched $U(T;\beta)$ estimates, on a logarithmic vertical scale.
    Each row corresponds to a different quenched instance. At this size, different RRG instances yield very similar TI integrands and reconstructed $Z(T;\beta)$ curves.
    }
    \label{fig:TI_RRG}
\end{figure*}

\begin{figure*}[ht!]
\includegraphics[]{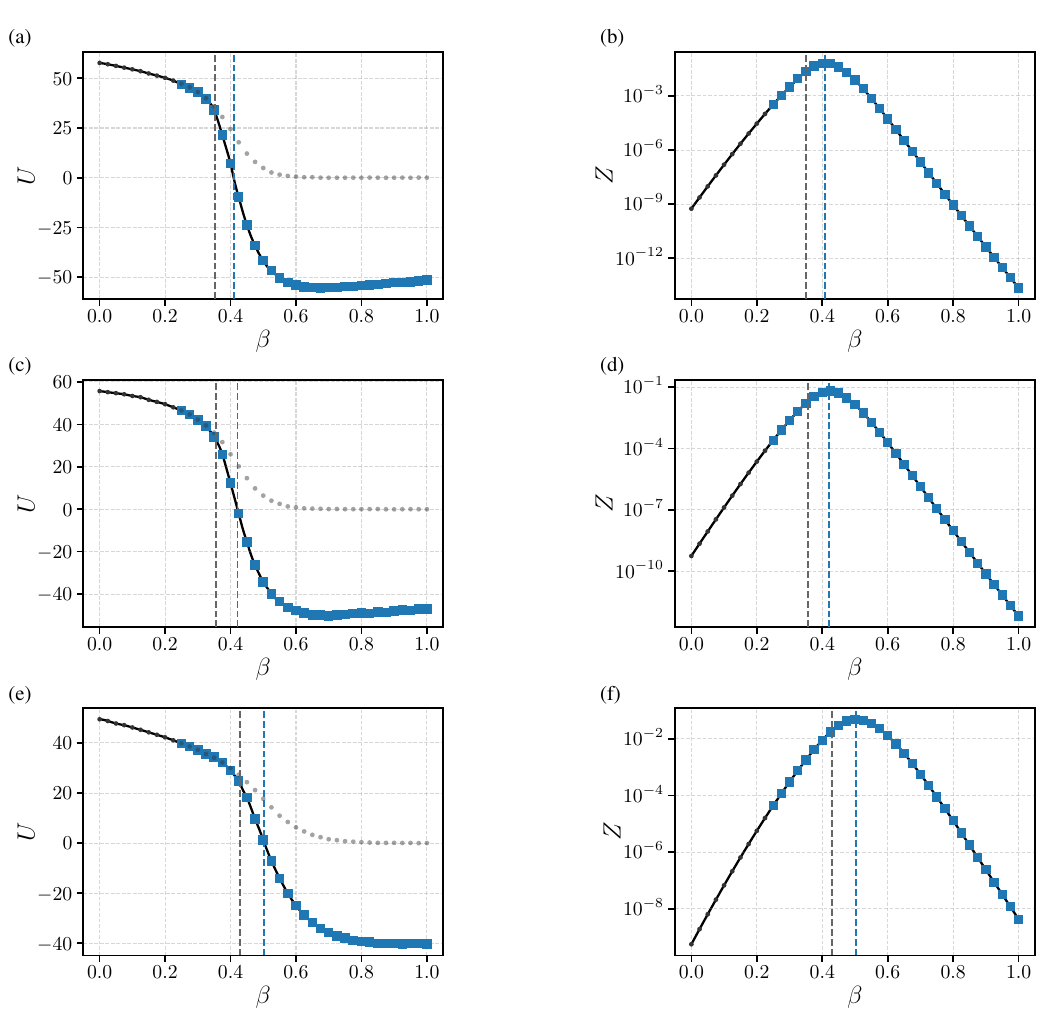}
    \vspace{-0.5cm}
    \caption{
    Representative thermodynamic-integration datasets for Erd\H{o}s--R\'enyi (ER, $c=3$) graphs at fixed size $N=100$ and trajectory length $T=50$.
    Panels \textbf{(a)}, \textbf{(c)}, and \textbf{(e)} show the TI integrand $U(T;\beta)=\partial_\beta \ln Z(T;\beta)$ for three different quenched instances; TPS estimates are shown as squares, while standard-MC estimates are shown as circles.
    Opaque circles denote MC points actually used in the matched integrand, whereas faded circles indicate MC measurements beyond the matching point, shown for reference only.
    Vertical dashed lines mark the graph-dependent scales $\beta_G$ (grey) and $\beta_M$ (blue).
    Panels \textbf{(b)}, \textbf{(d)}, and \textbf{(f)} show the corresponding $Z(T;\beta)$, reconstructed through thermodynamic integration of the matched $U(T;\beta)$ estimates, on a logarithmic vertical scale.
    Each row corresponds to a different quenched instance.
    In contrast with the RRG case (Fig.~\ref{fig:TI_RRG}), the dispersion across ER instances is substantial already at this size.
    }
    \label{fig:TI_ER}
\end{figure*}

To make explicit how the TI reconstruction is performed, we apply it to representative RRG and ER instances. These examples also highlight the different sample-to-sample fluctuations of the two ensembles, which motivate the rescaling discussed in Sec.~\ref{sec:ER}.
Figures~\ref{fig:TI_RRG} and \ref{fig:TI_ER} show representative RRG and ER graph instances, respectively, with one instance per row.

The left panels show the TI integrand
$U(T;\beta)=\partial_\beta \ln Z(T;\beta)$, i.e.\ the trajectory observable defined in Eq.~\eqref{eq:U_def} averaged over the appropriate trajectory ensemble; TPS and standard-MC estimates are denoted by squares and circles, respectively.
At sufficiently small $\beta$, configuration-space relaxation is fast on the chosen time scale and finite-time effects on the TPS estimate of $U$ become negligible within errors.
In this regime, we therefore use the standard MC estimator, which is computationally cheaper.
As already noted in Ref.~\cite{ZMW_2012}, for small enough $\beta$ and fixed $T$, relaxation is fast on the $T$ time scale, so that $U(T;\beta)$ becomes indistinguishable, within errors, from its $T\to\infty$ limit; this limit admits an equilibrium representation and can therefore be estimated efficiently by standard Monte Carlo.
We thus anchor the TI integrand at low $\beta$ with the standard-MC estimator and switch to TPS only beyond a matching point where the two estimators start to deviate appreciably.
In the figures, MC points actually used in the integration are shown as opaque circles, whereas MC measurements beyond the matching point are shown as faded circles for reference.

Vertical dashed lines in the left panels mark the two graph-dependent scales discussed in Sec.~\ref{sec:ER}: the topology-only estimate $\beta_G$ (grey) and the transition-maximum inverse temperature $\beta_M$ (blue).
The right panels show the corresponding transition probability $Z(T;\beta)$, reconstructed through thermodynamic integration of the matched $U(T;\beta)$ estimates, on a logarithmic vertical scale.
The maximum of $Z(T;\beta)$ along $\beta$, equivalently the zero of $U(T;\beta)$, provides the operational definition of $\beta_M$ used in Sec.~\ref{sec:ER}.

\subsection{Bootstrap error propagation for the TI integral}
\label{ssec:app_TI_errors}

\begin{figure*}[ht!]
\centering
\includegraphics[]{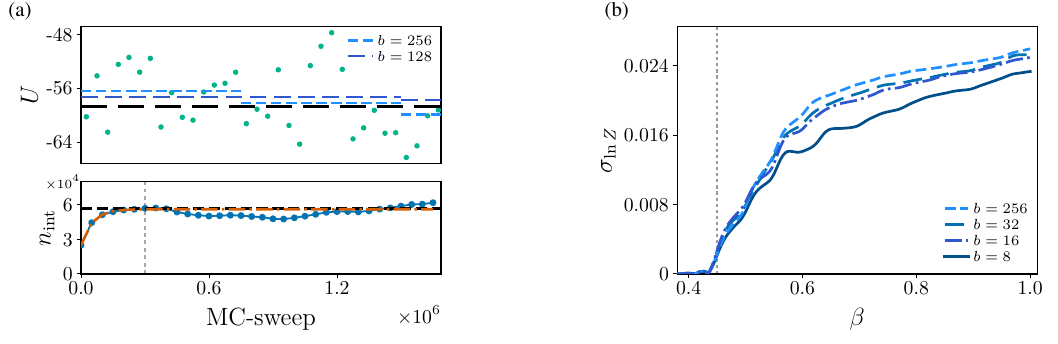}
\vspace{-0.6cm}
\caption{
Bootstrap diagnostic for the TI error on a representative RRG sample with \(c=3\), \(N=140\), at trajectory length \(T=90\).
(a) Upper subpanel: native-block estimates of \(U\) for the \(\beta=1\) PathsMC run, shown in a cropped window of a \(\sim2\times 10^8\)-sweep run; horizontal segments show the displayed \(b=256\) and \(b=128\) coarsenings, and the black line is the full-run average entering the TI curve.
Lower subpanel: cumulative integrated-autocorrelation diagnostic from the full native-block sequence; vertical guides mark the range used to read off \(n_{\rm int}\), and the dash-dotted reference overlaps the early-lag data, consistent with an effective single-exponential decay.
(b) Propagated \(\sigma_{\ln Z}\) along the corresponding TI curve for the blocking levels shown in the legend; the vertical guide marks the first PathsMC point entering the TI curve.
Additionally, blocking levels for \(b=128,64\) were included in the same analysis and fall within the displayed blocking-scale dependence, but are not shown for clarity.
}
\label{fig:TI_error_propagation}
\end{figure*}

In order to estimate the uncertainty in the transition probability \(Z(T;\beta)\) due to finite sampling of the TI integrand, we now apply a block-bootstrap analysis to the PathsMC estimates entering each TI curve. 
Since the TI factor multiplies the unnormalized dynamical estimate, the resulting dispersion enters additively in \(\ln k\).

For each PathsMC point of a TI curve, we start from the elementary block averages stored during the simulation, i.e.\ the finest available blocking of the MC-sweep history. 
These blocks are then grouped into a prescribed number \(b\) of coarser blocks. 
At each \(\beta\), bootstrap samples are obtained by resampling the \(b\) coarsened block averages with replacement, giving a bootstrapped value \(U^{(r)}(T;\beta)\) for each point of the \(\beta\)-ladder. 
For each bootstrap sample \(r\), only the values of \(U(T;\beta)\) are replaced by \(U^{(r)}(T;\beta)\); the \(\beta\)-grid, the selected TI points, and the interpolation/integration prescription used for the central curve are kept fixed. 
The resulting spread of \(\ln Z^{(r)}(T;\beta)\) defines \(\sigma_{\ln Z}(\beta)\). 
For this bootstrap check, the uncertainty propagation is restricted to the PathsMC points entering the matched TI curve; the standard-MC estimates used for the low-$\beta$ anchor have comparatively small statistical errors and are therefore held fixed. The first PathsMC points in the matching region are included in the propagation even when still compatible with the standard-MC estimates; the propagated TI uncertainty is therefore conservative in that region.

Figure~\ref{fig:TI_error_propagation} illustrates the diagnostic on a representative RRG sample with \(c=3\), \(N=140\), at trajectory length \(T=90\). 
Panel~(a) focuses on a \(\beta=1\) PathsMC run: it shows the native-block fluctuations of \(U\), the displayed \(b=256\) and \(b=128\) coarsenings, and the integrated-autocorrelation diagnostic. 
The decorrelation scale computed on the full run is already shorter than the block length corresponding to \(b=256\). 
In panel~(b), \(\sigma_{\ln Z}\) is propagated along the full TI curve for several values of \(b\). 
The resulting variation is moderate, indicating that the estimate is not driven by the precise blocking choice.

Across the representative curves checked, the propagated TI uncertainty remains both small and stable on the scale relevant to the subsequent analysis. 
We quote the diagnostic at the largest \(\beta\) of each TI curve, where the accumulated TI error is largest. 
For RRG, the case shown in Fig.~\ref{fig:TI_error_propagation} at \(\beta=1\) and three additional \(N=160\) checks give \(\sigma_{\ln Z}\simeq 0.02\text{--}0.04\). 
This corresponds to only a few percent in \(Z\), whereas the relevant changes of \(Z\) between sizes, and also the smaller graph-to-graph variations at fixed size, occur on much larger logarithmic scales. 
For three ER checks at \(N=200\), also around \(\beta=1\), the conservative estimates are larger but remain below \(0.08\) in \(\ln Z\), even for curves reaching \(Z\simeq 10^{-26}\); this is still small compared with the much larger sample-to-sample fluctuations at fixed \(N\) discussed in Sec.~\ref{sec:ER}. 
For the ZKC curve at \(T=40\) and \(\beta=0.9\), where longer PathsMC runs were needed to obtain clean \(\chi(t)\) curves for the fits across the different regimes, the same diagnostic gives \(\sigma_{\ln Z}\simeq 9\times10^{-3}\). 
The TI contribution to \(\ln Z\), and hence to \(\ln k\), is therefore too small to affect the rate and barrier analyses discussed in the main text.

\section{A three-state description in the 2D square-lattice Ising model}
\label{sec:app_2d_threestate}

\begin{figure*}[ht!]
\includegraphics[]{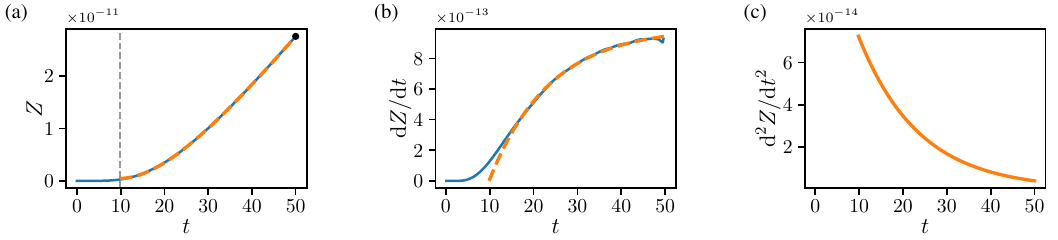}
\vspace{-0.5cm}
\caption{
2D square-lattice Ising model ($L=7$) at $\beta=1$: three-state fit used as a coarse-grained description in a setting where the crossover region has internal geometric structure.
\textbf{(a)} Transition probability $Z(t)$ and corresponding three-state fit (Eq.~\eqref{eq:3StateExSol}, with $t\to t-\tau_{\rm gran}$).
\textbf{(b)} Time derivative $dZ/dt$ from the same dataset, compared with the derivative of the fitted curve.
\textbf{(c)} Second derivative $d^2Z/dt^2$ of the fitted curve (not estimated directly from data, since numerical second derivatives strongly amplify noise).
}
\label{fig:app_2d_threestate}
\end{figure*}

In two dimensions, the transition pathway is not well represented by a single featureless intermediate state: the crossover region is instead shaped by geometric collective degrees of freedom associated with phase coexistence and interfacial configurations. As a result, a three-state reduction is not expected to capture in detail time-resolved observables such as $dZ/dt$. Nevertheless, fitting $Z(t)$ with a minimal three-state description remains informative as a coarse-grained baseline: it provides a compact characterization of the dominant timescales (onset and completion) and clarifies which features are plausibly tied to structured intermediate dynamics rather than to statistical noise or finite sampling.

We restrict attention to a single small lattice size $L=7$ ($N=L^2$) at a deep-ferromagnetic point, $\beta=1>\beta_c=\frac12\ln(1+\sqrt2)\simeq 0.4407$ (for $J=1$).
We analyze an aggregated dataset of $\sim 10^{10}$ transition trajectories obtained by combining $20$ independent runs at the same parameters.
Such aggregation is meaningful in this homogeneous setting because the transition is dominated by a single nucleation-and-growth scenario (up to lattice translations and interfacial shape fluctuations); pooling independent simulations therefore increases statistics without mixing qualitatively distinct classes of transition pathways.

We fit the measured $Z(t)$ following the same protocol used for ZKC in the three-state regime: we compare it to the three-state functional form Eq.~\eqref{eq:3StateExSol}, allowing for a short-time granular shift $\tau_{\rm gran}$, i.e.\ using $t\to (t-\tau_{\rm gran})$.
Concretely, we fit the kinetic parameters of Eq.~\eqref{eq:3StateExSol} (here the effective rate scale and the separation parameter $r$) together with $\tau_{\rm gran}$.
On the present aggregated dataset, fit-only uncertainties are small and are not reported.
For the present dataset, the fit yields $k\simeq 2.0\times 10^{-12}$, $r\simeq 2\times 10^{10}$, and $\tau_{\rm gran}\simeq 10$.
These numbers are consistent with the rate and transient scales discussed in Ref.~\cite{ZMW_2012}, once the different parametrizations are taken into account.

At fixed size, the three-state fit allows one to associate an effective interfacial barrier scale with the rate $k$ of entering the intermediate interfacial region,
without requiring any fit in $L$.
We define an effective interfacial barrier estimate $\Sigma_{\rm eff}$ by rewriting this rate as
\begin{equation}
k=t_0^{-1}e^{-2L\,\Sigma_{\rm eff}},
\end{equation}
where $t_0$ is a prefactor timescale characterizing the short-time dynamics.
Using the bare single-spin timescale of the continuous-time dynamics adopted here as reference, i.e.\ $t_0=1$ in our normalization of the rates (Sec.~\ref{ssec:setting}, Eq.~\ref{eq:rates}), one gets $\Sigma_{\rm eff}(1)\simeq 1.92$. The three-state fit also provides a nontrivial mesoscopic timescale, which can be estimated either through $\tau_{\rm gran}$ or through $\tau_I=(2rk)^{-1}$, yielding $\Sigma_{\rm eff}(\tau_{\rm gran})\simeq 1.76$ and $\Sigma_{\rm eff}(\tau_I)\simeq 1.74$; for the latter one has $\Sigma_{\rm eff}(\tau_I)=\ln(2r)/(2L)$. The proximity of these two values is consistent with the fact that, in the present 2D setting, $\tau_{\rm gran}$ and $\tau_I$ encode closely related short interfacial timescales. These values are much closer to the asymptotic Onsager value $\Sigma=1.7276$ than to the small-size regime heuristically characterized by $\Sigma\simeq 2$, in line with the crossover discussed in Ref.~\cite{ZMW_2012}. More generally, since the interfacial transient remains sub-exponential in $L$ whereas the transition rate is controlled by an $\exp(-2\Sigma L)$ scale, the estimates of $\Sigma_{\rm eff}$ obtained from the different choices of $t_0$ are expected to converge to the same large-$L$ limit.

Fig.~\ref{fig:app_2d_threestate} reports $Z(t)$, $dZ/dt$, and the fitted-curve second derivative $d^2Z/dt^2$ for this dataset.
The fit captures the global progress of the transition and reproduces the broad structure of $dZ/dt$, while panel~\textbf{(c)} makes explicit the curvature implied by this coarse-grained description.
At the same time, in 2D, the transition intermediate is geometrically structured by interface configurations, so the agreement should be read as an effective few-state description rather than as evidence for a single featureless intermediate state.

\section{Additional material on ZKC}
\label{sec:app_ZKCChiDerivatives}

\subsection{Low-\texorpdfstring{$\beta$}{beta} diagnostic: threshold sensitivity (\texorpdfstring{$M^\star=6$}{M*=6})}
\label{sec:app_ZKC_M6}

\begin{figure*}[ht!]
\includegraphics[]{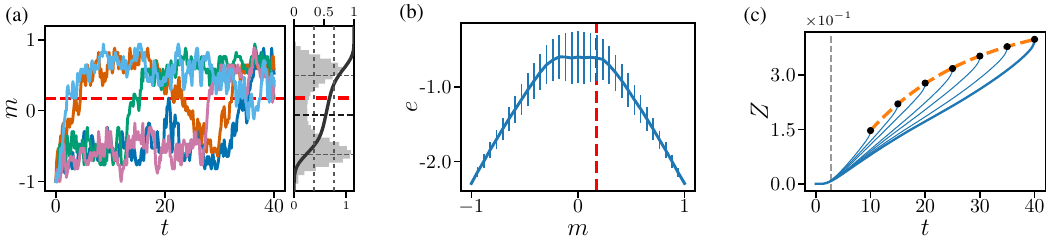}
\vspace{-0.6cm}
\caption{
Low-$\beta$ saturation regime on the ZKC graph at $\beta=0.25$, using the lower threshold $M^\star=6$ (same $\beta$ as in the main-text saturation example).
\textbf{(a)} Representative magnetisation trajectories $m(t)$ and density-normalised histogram of sampled occurrences.
\textbf{(b)} Average energy at fixed magnetisation and its dispersion; the red dashed line marks $m^\star=M^\star/N$.
\textbf{(c)} Thermodynamic--integration estimates at the sampled horizons $T$ (black markers), the corresponding reweighted family of curves $\{Z^{(M^\star=6)}(t)\}_T$ (blue solid curves), and the saturation fit~\eqref{eq:Zsatur_app} performed on the thermodynamic--integration points (yellow dashed curve).
Lowering $M^\star$ makes the basin indicator less sensitive to intra-basin dips below $m^\star$ while still effectively selecting the $(+)$ basin, thus providing a direct threshold-sensitivity check of the observed plateau.
}
\label{fig:ZKC_M6}
\end{figure*}

In the low-$\beta$ regime, the transition probability $Z(t)$ quickly reaches a plateau.
As discussed in Sec.~\ref{sec:ZKC}, the plateau value depends on the \emph{operational} definition of the final basin through the threshold $M^\star$ entering the indicator $\chi_{(+)}=\Theta(M-M^\star)$.
To assess the sensitivity to this choice, we repeat the analysis at the same $\beta$ using a lower threshold $M^\star=6$ (Fig.~\ref{fig:ZKC_M6}), which makes $\chi_{(+)}$ less sensitive to intra-basin magnetisation dips once the trajectory has already entered the $(+)$ well.

We fit the TI estimates with the same single-exponential saturation form,
\begin{multline}
\label{eq:Zsatur_app}
Z^{(M^\star)}_{\text{fit}}(t)=
c^{(M^\star)}
\Bigl[1-\exp\!\bigl(-k_{\rm sat}^{(M^\star)}\,(t-\tau_{\mathrm{gran}}^{(M^\star)})\bigr)\Bigr],
\\
\qquad t>\tau_{\mathrm{gran}}^{(M^\star)} .
\end{multline}
where $\tau_{\mathrm{gran}}^{(M^\star)}$ is an \emph{effective} horizontal shift capturing the early-time granularity of the thresholded observable, and should not be over-interpreted as a microscopic equilibration time: in practice the parameters
$\bigl(k_{\rm sat}^{(M^\star)},c^{(M^\star)},\tau_{\mathrm{gran}}^{(M^\star)}\bigr)$ are correlated in the fit.

For $M^\star=6$ we obtain
$k_{\rm sat}^{(M^\star=6)}=0.052\pm0.002$,
$c^{(M^\star=6)}=0.464\pm0.004$,
and
$\tau_{\mathrm{gran}}^{(M^\star=6)}=2.7\pm0.2$.
The key point is the interpretation of $c^{(M^\star)}$.
By definition, $c^{(M^\star)}=\lim_{t\to\infty}Z^{(M^\star)}(t)$ is the asymptotic probability of satisfying the \emph{threshold condition} $M>M^\star$, not the full equilibrium weight of the $(+)$ basin.
In a strictly $\mathbb{Z}_2$-symmetric situation at zero external field, the two ferromagnetic wells carry equal equilibrium weight $1/2$.
Therefore, if the threshold is low enough that typical fluctuations inside $(\pm)$ do \emph{not} spend appreciable time in the central region $|M|\le M^\star$, then $c^{(M^\star)}$ should approach $1/2$.
Conversely, any non-negligible equilibrium weight in the threshold-excluded region $|M|\le M^\star$ reduces $c^{(M^\star)}$ below $1/2$.

Under these assumptions, the quantity
\begin{equation}
\label{eq:mass_outside_basins}
\frac12-c^{(M^\star)}
\end{equation}
provides an estimate of the equilibrium probability mass \emph{per basin side} that lies outside the corresponding threshold-defined well, i.e.\ in the central region $|M|\le M^\star$.
Equivalently, the total equilibrium weight in the threshold-excluded central region is
$P_{\rm eq}(|M|\le M^\star)=1-2c^{(M^\star)}$.
For $M^\star=6$ we find
\begin{equation}
\label{eq:mass_outside_basins_num}
\frac12-c^{(M^\star=6)} = 0.036\pm0.004,
\end{equation}
supporting an effectively two-well picture in this low-$\beta$ regime and confirming that the deviation of $c^{(M^\star=20)}$ from $1/2$ is largely a threshold effect.

\subsection{Derivatives and transition-time statistics}
\label{ssec:app_transition_times}

\begin{figure*}[ht!]
\includegraphics[]{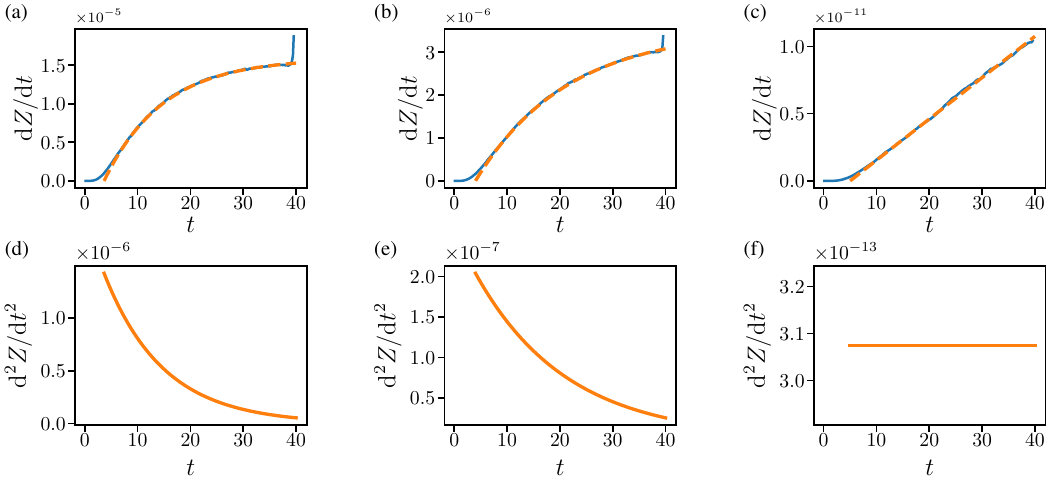}
\vspace{-0.5cm}
\caption{
\textbf{(a)--(c)} First derivative $dZ/dt$.
Blue: central finite-difference derivative of the reconstructed $Z(t)$ curve.
Orange: analytical derivative of the fitted form used in Sec.~\ref{sec:ZKC}
(three-state at $\beta=0.50,0.55$, quadratic onset at $\beta=0.90$).
Panels \textbf{(a)}, \textbf{(b)}, and \textbf{(c)} correspond respectively to $\beta=0.50$, $0.55$, and $0.90$; for the corresponding $Z(t)$ curves and fits at $\beta=0.55$ and $0.90$, see Fig.~\ref{fig:ZKC_regimes}(g) and Fig.~\ref{fig:ZKC_regimes}(j).
A small end-of-window feature close to $t\simeq T$ is an operational endpoint effect induced by the threshold-based basin definition ($M^\star,m^\star$) and is discussed in Appendix~\ref{sec:app_moderateBeta}.
\textbf{(d)--(f)} Second derivative $d^2Z/dt^2$ of the fitted curves only, for the same three values of $\beta$.
In the persistence regime and within the short-time window where
$p_{\uparrow}(t)\simeq\tilde k$, the curvature provides an operational proxy
for the completion-time density $p_{\rightsquigarrow}(t)$ implied by the fit;
estimating $d^2Z/dt^2$ directly from reconstructed data is unstable because differentiation amplifies finite-sampling fluctuations and endpoint effects.
}
\label{fig:app_ZKC_derivatives}
\end{figure*}

We discuss how time derivatives of the transition probability encode
information on \emph{first-entry} and \emph{completion} time scales.
Throughout this subsection we assume a \emph{persistence} condition:
once the trajectory has entered the $(+)$ basin (according to the chosen
operational indicator), the probability of leaving it again within the time window of interest is negligible.
Under this assumption, the transition probability can be written as
\begin{equation}
  Z_{(+)}(t)=\int_0^t dt'\,p_{\rm in}(t'),
\end{equation}
where $p_{\rm in}(t)$ is the p.d.f.\ of the \emph{first entry time} into $(+)$.
Equivalently,
\begin{equation}
  \frac{d}{dt}Z_{(+)}(t)=p_{\rm in}(t).
  \label{eq:dZ_is_entrypdf}
\end{equation}

A useful refinement is to decompose the overall transition into two
sub-processes:
(i)~an \emph{escape trigger} that takes the system out of the $(-)$ basin,
and (ii)~a \emph{completion dynamics} that carries it from the first escape
to the operational entry in $(+)$.
Let $T_{\uparrow}$ be the (random) time at which the escape trigger occurs,
with p.d.f.\ $p_{\uparrow}(t)$, and let $\Delta$ be the (random) completion
time, with p.d.f.\ $p_{\rightsquigarrow}(\delta)$.
Under stationary conditions, $p_{\rightsquigarrow}$ depends only on the
duration $\delta$ (not on the absolute start time).
The first-entry time is then $T_{\rm in}=T_{\uparrow}+\Delta$, so that
\begin{equation}
  p_{\rm in}(t)
  =\int_0^t dt''\,p_{\uparrow}(t'')\,p_{\rightsquigarrow}(t-t''),
  \label{eq:Pin_convolution}
\end{equation}
and, combining with~\eqref{eq:dZ_is_entrypdf},
\begin{equation}
  \frac{d}{dt}Z_{(+)}(t)
  =\int_0^t dt''\,p_{\uparrow}(t'')\,p_{\rightsquigarrow}(t-t'').
  \label{eq:dZ_convolution}
\end{equation}

If, in addition, equilibration within the $(-)$ basin is fast on the
time scales we probe, the escape triggering can be treated as a stationary
Poisson process with rate $\tilde k$. In that case, the triggering time
has an exponential waiting-time density,
\begin{equation}
  p_{\uparrow}(t)=\tilde k\,e^{-\tilde k t}.
\end{equation}
In the time window relevant for our analysis, we focus on times short compared to the mean triggering time, $t\ll \tilde k^{-1}$, so that one may linearise
the exponential and approximate
\begin{equation}
  p_{\uparrow}(t)\simeq \tilde k .
  \label{eq:Pu_constant_approx}
\end{equation}
Substituting Eq.~\eqref{eq:Pu_constant_approx} into Eq.~\eqref{eq:dZ_convolution}
yields
\begin{equation}
  \frac{d}{dt}Z_{(+)}(t)
  \simeq
  \tilde k \int_0^t d\delta\,p_{\rightsquigarrow}(\delta)
  \;=\;\tilde k\,F_{\rightsquigarrow}(t),
  \label{eq:dZ_kF}
\end{equation}
where $F_{\rightsquigarrow}(t)=\int_0^t d\delta\,p_{\rightsquigarrow}(\delta)$
is the cumulative distribution of completion times.

Equation~\eqref{eq:dZ_kF} provides a direct operational interpretation of the
shape of $Z_{(+)}(t)$.
If $p_{\rightsquigarrow}$ has (effective) support on $[0,\delta_{\max}]$,
then for $t\gtrsim\delta_{\max}$ one has $F_{\rightsquigarrow}(t)\simeq 1$
and therefore
\begin{equation}
  \frac{d}{dt}Z_{(+)}(t)\simeq \tilde k,
\end{equation}
so that $Z_{(+)}(t)$ becomes linearly growing after the completion-time transient.
Conversely, any curvature of $Z_{(+)}(t)$ at early/intermediate times directly
reflects the build-up of $F_{\rightsquigarrow}(t)$, i.e.\ the statistics of
completion times encoded in $p_{\rightsquigarrow}$.

We include $\beta=0.50$ as an additional representative point within the
three-state window (not shown in Fig.~\ref{fig:ZKC_regimes}) to validate the
derivative-level consistency of the three-state fit.

In our numerical implementation, $Z(t)$ is reconstructed from a single long-$T$ TI run via the reweighting identity described in Sec.~\ref{ssec:rateTrick}.
To compare derivatives consistently with the fitted functional forms, we apply the same microscopic time shift $\tau_{\mathrm{gran}}$ used in the fits (Sec.~\ref{sec:ZKC}), i.e.\ we interpret it as an operational offset setting the effective time origin of the basin indicator.

Figure~\ref{fig:app_ZKC_derivatives} compares the central finite-difference
estimate of $dZ/dt$ computed from the reconstructed $Z(t)$ (blue) with the
analytical derivative of the fitted forms (orange). The agreement is excellent
in the three-state regime ($\beta=0.50,0.55$) and in the quadratic-onset regime
($\beta=0.90$); the corresponding $Z(t)$ curves and fits for $\beta=0.55$ and
$\beta=0.90$ are shown in Fig.~\ref{fig:ZKC_regimes}(g) and Fig.~\ref{fig:ZKC_regimes}(j).

Moreover, as the trigger can be treated as approximately constant, $p_{\uparrow}(t)\simeq\tilde k$, Eq.~\eqref{eq:dZ_kF} implies
\begin{equation}
  \frac{d^2}{dt^2}Z_{(+)}(t)\simeq \tilde k\,p_{\rightsquigarrow}(t),
\end{equation}
so that the curvature of $Z_{(+)}(t)$ provides a direct operational proxy for the distribution of completion times. In practice, estimating $d^2Z/dt^2$ from the reconstructed $Z(t)$ is unstable because differentiation strongly amplifies finite-sampling fluctuations and endpoint irregularities; we therefore report $d^2Z/dt^2$ for the fitted curves only (bottom row), as a compact diagnostic of the completion-time profile implied by the fit.

\begin{figure}[ht!]
  \includegraphics{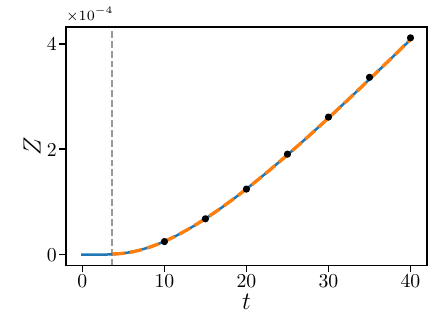}
  \vspace{-0.25cm}
    \caption{
    Transition-probability curve $Z(t)$ for ZKC at $\beta=0.50$ (not shown in Fig.~\ref{fig:ZKC_regimes}),
    reconstructed from the single long-$T$ TI run as in Sec.~\ref{ssec:rateTrick}.
    Black markers are the corresponding discrete TI estimates.
    The dashed curve is the three-state fit used in Sec.~\ref{ssec:app_transition_times}
    and in Fig.~\ref{fig:app_ZKC_derivatives}.
    }
  \label{fig:ZKC_Z_b050_appendix}
\end{figure}

The corresponding $Z(t)$ curve and fit at $\beta=0.50$ are shown in
Fig.~\ref{fig:ZKC_Z_b050_appendix}. The best-fit parameters are
$r=1708\pm 11$, $k=(3.06\pm 0.03)\times 10^{-5}$,
$\tau_{\mathrm{gran}}=4.10\pm0.03$, and $C=(0\pm2)\times 10^{-5}$.

\section{Not-too-low temperatures: threshold recrossings and finite-size effects}
\label{sec:app_moderateBeta}

At not-too-low inverse temperatures, the ferromagnetic wells broaden and the threshold $m^\star$ used to define
$\chi_{(+)}(t)=\Theta(m(t)-m^\star)$ ceases to be a persistent indicator of basin occupancy.
Even after a genuine $(-)\!\to\!(+)$ transition has occurred, intra-basin fluctuations can bring the magnetisation temporarily below $m^\star$,
producing repeated threshold recrossings without an actual return to the $(-)$ well.
This affects the practical interpretation of the time-dependent observable that we fit to extract rates, namely
\begin{equation}
  \bar\chi(t)\equiv
  \langle \chi_{(+)}(t)\rangle_{X,Y,T,\beta} .
\end{equation}
In the deep low-temperature regime, one typically has $\chi_{(+)}(t)$ coinciding with the persistent basin label (up to the single jump), so that $Z_{X,Y}(T)\bar\chi(t)$ directly measures, within the persistence regime, the occupancy probability of the $(+)$ basin.
When recrossings are present, $\bar\chi(t)$ still displays a linear-growth window, but its slope must be interpreted with care.

\begin{figure*}[ht!]
\includegraphics[]{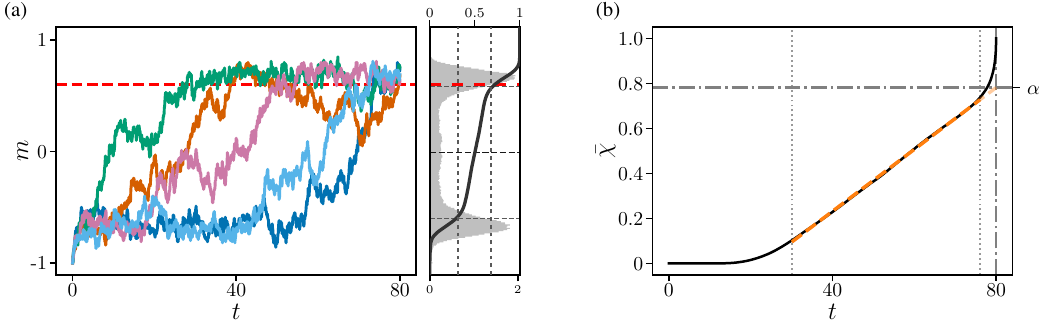}
    \vspace{-0.7cm}
    \caption{
    Example of threshold recrossings at not-too-low $\beta$.
    \textbf{(a)} Magnetisation trajectories for an ER instance with $N=200$ at $\beta=0.525$, using the same threshold value as in Sec.~\ref{sec:ER}, namely $m^\star=0.6$, marked with the red dashed horizontal line.
    After the transition, $m(t)$ repeatedly crosses the threshold $m^\star$, producing intermittent switching of $\chi_{(+)}(t)=\Theta(m(t)-m^\star)$ without a return to the opposite basin.
    \textbf{(b)} Corresponding $\bar\chi(t)=\langle\chi_{(+)}(t)\rangle$.
    The dashed orange line is a fit over the linear-growth window (delimited by the vertical dotted grey lines), whose slope defines $s_g$, with $\tilde k_g=Z_{X,Y}(T)s_g$.
    Extrapolating the same fit to $t=T$ yields $\bar\chi_{\rm lin}(T)\equiv \alpha_g<1$.
    The upward hook close to $t=T$ is induced by the hard final constraint at time $T$.
    }
    \label{fig:app_moderateBeta}
\end{figure*}

A convenient description is obtained by separating slow inter-basin dynamics from fast intra-basin fluctuations.
Let $S(t)\in\{-,+\}$ denote a coarse-grained basin label, and assume that, conditioned on $S(t)=+$, the threshold variable
$\chi_{(+)}(t)$ rapidly equilibrates on a time scale $\tau_{\rm rec}$ to a constant fraction
\begin{equation}
  \alpha_g(\beta)\equiv \mathbb P\!\left(\chi_{(+)}=1 \,\big|\, S=+,\, g\right), \qquad 0<\alpha_g\le 1,
\end{equation}
where we explicitly allow for graph-to-graph fluctuations (at fixed $\beta$).
Under a clear separation of time scales,
\begin{equation}
  \tau_{\rm rec}\ll T \ll k_g^{-1},
\end{equation}
one expects, throughout the post-transient linear-growth regime, where threshold fluctuations at $t$ and $T$ are effectively decorrelated,
\begin{equation}
  Z_{X,Y}(T)\bar\chi(t)
  \simeq
  \alpha_g^2(\beta)
  \,\mathbb P\!\left(S(t)=+ \,\big|\, g\right).
\end{equation}
Since $\mathbb P(S(t)=+\,|\,g)$ grows linearly with $t$ in the activated regime (after the transient time $\tau$), the post-transient tail is preserved but its slope is renormalised:
\begin{equation}
  Z_{X,Y}(T)\bar\chi(t)
  \simeq
  \tilde k_g\,(t-\tau),\qquad
  \tilde k_g=
  \alpha_g^2(\beta)
  \,k_g.
\end{equation}
Therefore, in the presence of threshold recrossings, the slope of the linear-growth window, after multiplication by $Z_{X,Y}(T)$,
yields an effective rate $\tilde k_g$.
Operationally, by construction one has $\chi_{(+)}(T)=1$ on every sampled trajectory, hence $\bar\chi(T)=1$ irrespective of $\alpha_g$:
this produces an endpoint distortion close to $t=T$, the final ``hook'' in Fig.~\ref{fig:app_moderateBeta}(b).
We therefore fit $\bar\chi(t)$ only within the intermediate linear-growth window, excluding the endpoint region where the hook is visible:
the slope of this fit defines $s_g$, with $\tilde k_g=Z_{X,Y}(T)s_g$,
while its extrapolated value at $t=T$ provides $\alpha_g(\beta)=\bar\chi_{\rm lin}(T)$.

The corrected basin-to-basin rate is then obtained as
\begin{equation}
  k_g=
  \tilde k_g/\alpha_g^2
  =
  \frac{Z_{X,Y}(T)s_g}{\alpha_g^2}.
\end{equation}
At low temperatures, one finds $\alpha_g\simeq 1$ and the correction becomes negligible.

Fig.~\ref{fig:app_moderateBeta} shows a representative case where the effect is manifest:
magnetisation traces display frequent excursions below $m^\star$ even after the transition,
while $\bar\chi(t)$ still exhibits a well-defined intermediate linear-growth regime followed by the endpoint hook.
In this regime, extracting $s_g$ from the linear window, combining it with the absolute normalisation $Z_{X,Y}(T)$, and correcting by the factor $\alpha_g^2$ obtained from the same extrapolation yields a stable estimate of the basin-to-basin rate.

\begin{figure*}[ht!]
\includegraphics[]{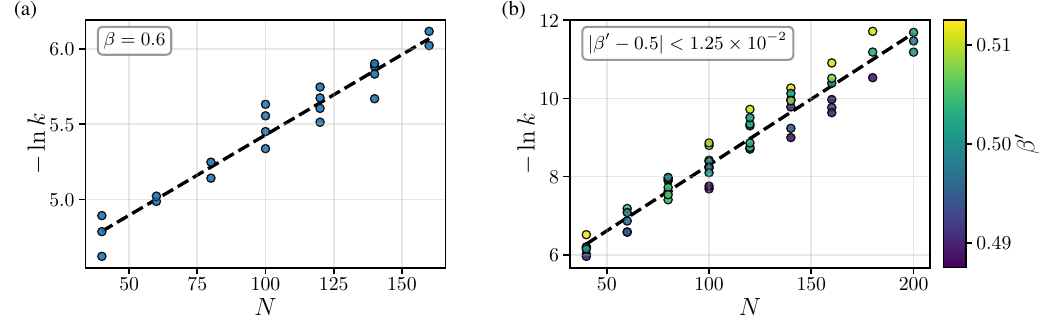}
    \vspace{-0.7cm}
  \caption{
    Arrhenius plot at not-too-low inverse temperature, using the same models and extraction protocols as in the main text.
    \textbf{(a)} Random regular graphs (RRG, $c=3$) at $\beta=0.6$.
    \textbf{(b)} Erd\H{o}s--R\'enyi graphs (ER, $c=3$) at fixed rescaled inverse temperature $\beta'=0.5$ (binning width $\epsilon=0.025$), using the $\beta_M$-based rescaling of Sec.~\ref{sec:ER} (i.e., $\beta_g=\beta_M(g)$); colors encode the rescaled inverse temperature $\beta'$ of datapoints before binning.
    Markers show $-\ln k_g$ versus $N$ (one marker per graph instance), with $k_g=\tilde k_g/\alpha_g^2$ corrected for threshold recrossings as discussed above.
    The dashed line is a linear fit over the available sizes.
  }
\label{fig:app_ArrheniusNotTooLow}
\end{figure*}

Beyond correcting the rate estimator itself, it is useful to check whether the resulting size scaling remains compatible with the leading Arrhenius picture in this regime.
As a diagnostic, we report in Fig.~\ref{fig:app_ArrheniusNotTooLow} Arrhenius plots of the corrected transition rates, $k_g=\tilde k_g/\alpha_g^2$, at representative inverse temperatures.
Panel~\textbf{(a)} shows the RRG case ($c=3$, same $m^\star$ and boundary conditions as in Sec.~\ref{sec:RRG}) at $\beta=0.6$.
Panel~\textbf{(b)} shows the analogous ER diagnostic at fixed \emph{rescaled} inverse temperature $\beta'=\tilde\beta$ (here $\tilde\beta=0.5$), using the $\beta_M$-based temperature rescaling and the same binning protocol in $\beta'$ as in Sec.~\ref{sec:ER}.
In both cases, the corrected rates remain compatible with a dominant linear dependence on $N$, although in this low-barrier regime prefactor contributions are comparatively more important and can result in residual scatter or weak departures from linearity.

\section{ER rescaling with the topology-only scale \texorpdfstring{$\beta_G$}{betaG}}
\label{sec:app_betaG_rescaling}

\begin{figure*}[ht!]
\includegraphics[]{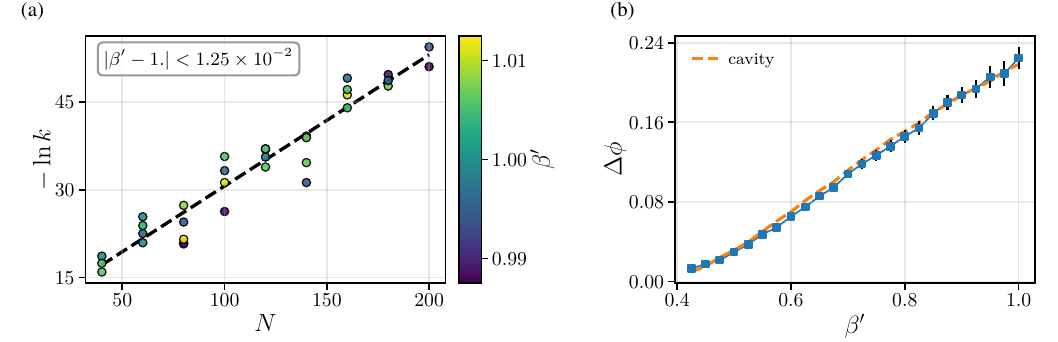}
\vspace{-0.55cm}
\caption{
Erd\H{o}s--R\'enyi graphs with average connectivity $c=3$: rescaling based on the topology-only scale $\beta_G(g)$ (Eq.~\eqref{eq:betaG_def}), shown for comparison with the $\beta_M$-based analysis of Fig.~\ref{fig:er_rateExtraction}.
\textbf{(a)} $-\ln k$ versus $N$ at rescaled inverse temperature $\beta'=1$, obtained by mapping each instance through Eq.~\eqref{eq:newBetas_betaG} and assigning datapoints to bins of width $\epsilon=0.025$ (same procedure as in Sec.~\ref{sec:ER}; compare with Fig.~\ref{fig:er_rateExtraction}(b)).
\textbf{(b)} Barrier curve $\Delta\phi(\beta')$ extracted from the slopes of the corresponding linear fits in $N$ for each $\beta'$-bin (same procedure as for Fig.~\ref{fig:er_rateExtraction}(c)); the dashed line is the Bethe (cavity/BP) prediction for the ER ensemble computed by population dynamics (as described in Sec.~\ref{sec:ER}).
}
\label{fig:ER_betaG_rescaling}
\end{figure*}

As a robustness check of the instance-dependent rescaling introduced in Sec.~\ref{sec:ER}, we repeat the ER analysis using the topology-only inverse-temperature scale $\beta_G(g)$ in Eq.~\eqref{eq:betaG_def}, instead of the operational scale $\beta_M(g)$ extracted from thermodynamic integration.
The aim is to verify that the collapse of transition rates is not tied specifically to the proxy $\beta_M$, and to compare the resulting barrier estimates with the static Bethe (cavity/BP) prediction discussed in Sec.~\ref{sec:ER} and reported in Fig.~\ref{fig:er_rateExtraction}(c).

We follow the same pipeline as in Sec.~\ref{sec:ER}---in particular, we use the same set of retained ER instances and the same binning protocol---changing only the definition of the rescaled inverse temperature. The barrier is then extracted from the same linear-fit procedure applied to $-\ln k$ versus $N$ in the main text.

Concretely, for each ER instance $g$ retained in the main analysis (i.e.\ satisfying the criterion $T>\tau(\beta=1)$ of Sec.~\ref{sec:ER}), we define
\begin{equation}
  \beta' \equiv \beta\,\frac{\overline{\beta_G}}{\beta_G(g)} ,
  \label{eq:newBetas_betaG}
\end{equation}
where $\overline{\beta_G}\equiv \mathrm{median}\{\beta_G(g)\}$ is taken over the same set of retained instances (each graph counted once), consistently with the definition of $\overline{\beta_M}$ used in Eq.~\eqref{eq:newBetas}.
As in Sec.~\ref{sec:ER}, the mapping in Eq.~\eqref{eq:newBetas_betaG} sends the common simulation ladder in $\beta$ to instance-dependent values of $\beta'$, so we bin the resulting measurements on a uniform grid $\{\tilde\beta_n=n\epsilon\}_{n\in\mathbb{N}_0}$ with the same bin width $\epsilon=0.025$ used in the main text.

Fig.~\ref{fig:ER_betaG_rescaling} mirrors Fig.~\ref{fig:er_rateExtraction} to illustrate the $\beta_G$-based rescaling.
Panel (a) shows $-\ln k$ versus $N$ in the bin centered at $\beta'=1$, similarly to Fig.~\ref{fig:er_rateExtraction}(b), and panel (b) reports the corresponding barrier curve $\Delta\phi(\beta')$ extracted by repeating the binning-and-fit procedure for each $\beta'$-bin, similarly to Fig.~\ref{fig:er_rateExtraction}(c).
The same Bethe (cavity/BP) prediction for the ER ensemble, computed by population dynamics as described in Sec.~\ref{sec:ER}, is shown as a dashed line in panel (b).
Overall, the $\beta_G$-based rescaling yields a collapse and barrier estimates compatible with the $\beta_M$-based analysis within the uncertainties over the accessible sizes, confirming that the ER conclusions of Sec.~\ref{sec:ER} are robust to the choice of proxy scale.
No systematic deviations between the two prescriptions are visible at the accessible sizes.
This is consistent with $\beta_G$ targeting the thermodynamic temperature scale primarily, whereas $\beta_M$---being extracted from finite-$T$ transition data---may partly absorb subleading finite-size/kinetic contributions and thus optimise the collapse of finite-$N$ dynamical rates (cf.\ the discussion following Eq.~\eqref{eq:betaCorr}).
As expected from the slightly weaker collapse compared to the $\beta_M$ prescription, the statistical uncertainty on the fitted slopes (and thus on $\Delta\phi$) increases mildly at the largest rescaled inverse temperatures.

\end{document}